\documentclass{article}
\usepackage[thinc]{esdiff}
\usepackage{lipsum}
\usepackage{amsmath,amssymb,amstext, amsthm}
\usepackage[style=authoryear, backend=biber, natbib=true]{biblatex}
\usepackage{cancel}
\usepackage{xcolor}
\usepackage{smaller}
\usepackage{mathtools}
\usepackage[colorlinks=true, linkcolor=blue, citecolor=blue, urlcolor=blue]{hyperref}
\usepackage[a4paper,margin=1.5cm,columnsep=1cm]{geometry}

\usepackage{authblk}
\usepackage{subcaption}
\usepackage{siunitx}
\usepackage{hyperref}

\theoremstyle{plain}

\newtheorem{rem}{Remark}[section]

\newcommand{\dif}{\mathrm{d}}

\newcommand{\vort}[1]{\mathbf{k} \cdot \nabla \times \mathbf{#1}}

\renewcommand{\thesubsubsection}{(\roman{subsubsection})} 

\usepackage{titlesec}
\titleformat{\subsection}{\bfseries\normalsize}{\thesubsection}{1em}{}
\titleformat{\subsubsection}{\normalfont\normalsize}{\thesubsubsection}{1em}{}

\renewcommand{\thesubsection}{\thesection.\alph{subsection}} 
\usepackage{hyperref}

\newcommand\blfootnote[1]{%
  \begingroup
  \renewcommand\thefootnote{}\footnote{#1}%
  \addtocounter{footnote}{-1}%
  \endgroup
}

\title{\textbf{\Large{Access to the quasi-geostrophic model from within the BK19 compressible flow solver}}}
\author[1]{Amir Maghoul}
\affil{FB Mathematik \& Informatik, Freie Universität Berlin; Arnimallee 2-9, 14195 Berlin}
\author[1]{Rupert Klein}
\date{\smaller November 2025}

\begin{document}

\maketitle
\blfootnote{Corresponding author: Amir Maghoul \textit{amir.maghoul@fu-berlin.de}}

\begin{abstract}
    We extend BK19 the semi-implicit finite-volume solver, to provide access to the quasi-geostrophic (QG) regime within the same compressible framework that already admits soundproof and hydrostatic limits. In order to achieve this goal, we incorporate the asymptotic framework from the continuous theory into the numerical scheme. Adopting the usual asymptotic setting for the quasi-geostrophic regime and using it directly in the numerical model, we perform a scale analysis of the compressible Euler equations in the synoptic length and time scales in the limit of small Rossby number. In this limit, the discrete implicit step recovers geostrophic balance, together with the pseudo-incompressible divergence constraint. We construct a staggered-grid curl operator consistent with the discrete derivatives and use it to define the vorticity in the numerical framework. An asymptotic analysis of the fully discrete scheme then yields a Poisson-type equation for the leading-order Exner-pressure perturbation that coincides with the QG pressure Poisson equation in the continuous theory. As a result of this analysis, we introduce two additional blending switches that (i) convert the elliptic pressure problem from the full Helmholtz to its QG Poisson-type equation and (ii) control whether the horizontal momenta are updated prognostically or diagnosed from geostrophic balance. In combination with the existing soundproof and hydrostatic switches, the resulting algorithm can be operated in fully compressible, pseudo-incompressible, hydrostatic, and QG modes without changing the underlying discretization. The functionality of the numerical scheme in the QG regime is then tested by two idealized vortex test cases to show how stable the vortices remain in the QG run and that, as the Rossby number is reduced, the full compressible scheme converges to slow, balanced QG dynamics.
\end{abstract}



\section{Introduction}\label{sec: introduction}

\subsection{Motivation: Extending blended multimodel solvers to the quasi-geostrophic regime}

Reduced models play an important role in theories for atmospheric dynamics. By utilizing scale analysis and asymptotic limits of the compressible Euler equations, one derives simplified forms of the equations associated with particular flow regimes that are compatible with the leading-order balances  \citep{Pedlosky1987GeophysicalDynamics,Klein2010Scale-dependentFlows}. In this way, they distinguish between processes that are dynamically essential and those that are negligible or insignificant at the scales of interest. A few important reduced models are the anelastic and pseudo-incompressible systems, the quasi-geostrophic and semi-geostrophic approximations, and the hydrostatic primitive equations \citep{Hoskin1972AtmosphericSolution,Lipps1982ACalculations.,Durran1989ImprovingApproximation,Pedlosky1987GeophysicalDynamics,Bannon1996OnAtmosphere,Cullen2003TheMeteorology,Klein2010Scale-dependentFlows}. Together, these models, constructed in the limits of vanishing Mach, Rossby, and Froude numbers provide a structured understanding of the different characteristics of atmospheric flows. 

From a numerical point of view, these asymptotic limits are not just conceptual. \citet{Cullen2007ModellingFlows} argued that discretizations of the compressible solvers for the atmospheric flows should converge towards the appropriate reduced models as the relevant parameters approach their limiting values. Numerical schemes with this property are often referred to as \emph{asymptotic-preserving} or \emph{asymptotically adaptive} \citep{Jin1999EfficientEquations, Jin2010AsymptoticReview, Filbet2010ASources, Jin2022Asymptotic-PreservingProblems, Noelle2014ADynamics, Bispen2017AsymptoticGravitation,Klein2001AsymptoticMechanics,Benacchio2014ADynamics}. When blending parameters are also introduced, allowing the same numerical formulation to function in either fully compressible mode or in one (or more) asymptotic regimes, a single dynamical core can be used to study different flow regimes without changing the underlying discretization and code structure \citep{Benacchio2014ADynamics,Klein2016ADynamics,Benacchio2016Semi-implicitPerspective, Hastermann2021BalancedSystems, Chew2022ATests}. In these blended -- or \emph{seamless} -- formulations, the reduced models are recovered as limiting cases of the underlying compressible scheme, rather than as separate stand-alone discretizations.

\citet{Benacchio2019ADynamics} (hereafter BK19) presented a semi-implicit finite-volume scheme for the rotating compressible Euler equations formulated in terms of mass, momentum, and mass-weighted potential temperature. By utilizing the structural relationship between the compressible, pseudo-incompressible, and hydrostatic formulations, BK19 introduced blending parameters that provide \emph{seamless access} in the sense of the last paragraph to soundproof and hydrostatic dynamics within a unified numerical framework \citep{Klein2009AsymptoticsEquations,Klein2010RegimeModels,Benacchio2014ADynamics, Hastermann2021BalancedSystems, Chew2022ATests}. The resulting algorithm, also referred to as BK19 algorithm in the present work, achieves stability for large time-steps which are essentially governed by advective speeds and  provide a blended multimodel solver with soundproof and hydrostatic switches \citep{Klein2016ADynamics}.

A major remaining gap in this program concerns the synoptic-scale midlatitude dynamics that are classically described by quasigeostrophic (QG) theory \citep{Vallis2017AtmosphericEdition, Gill1982Atmosphere-oceanDynamics,Pedlosky1987GeophysicalDynamics,Zeytounian1990AsymptoticFlows}. In BK19, the authors deferred the treatment of the quasi-geostrophic regime to future work. The QG model provides a paradigmatic framework for large-scale, slowly evolving, balanced motions in which both acoustic and internal gravity waves are asymptotically filtered out and the flow is governed by potential-vorticity (PV) advection on the advective timescale \citep{Pedlosky1987GeophysicalDynamics,Vallis2017AtmosphericEdition}. Embedding such a QG limit consistently into BK19 would extend the range of accessible asymptotic regimes to those of primary relevance for midlatitude weather systems and climate variability.

The present paper addresses this point. Building on the scale-dependent asymptotic framework of \citet{Klein2010Scale-dependentFlows} as formulated in \citet{Klein2022QG-DL-Ekman:Framework} for the QG regime, we derive and implement a quasi-geostrophic limiting regime within the BK19 discretization. Rather than viewing QG solely as a stand-alone PV model, we use the systematic asymptotic approach to characterize QG as part of a hierarchy of soundproof, hydrostatic, and geostrophically balanced models, all expressed in the same Exner-pressure, potential-temperature, and momentum variables. This perspective allows us to introduce a \emph{geostrophic update switch} together with a \emph{QG Poisson-equation switch} into BK19, consistent with the underlying compressible formulation and with the existing soundproof and hydrostatic switches. This implements the targeted seamless access to QG dynamics within the BK19 scheme.

\subsection{Classical versus modern quasi-geostrophic frameworks}

The classical quasigeostrophic system is derived from the hydrostatic, rotating Boussinesq or primitive equations under the assumption of small Rossby, Mach, and Froude numbers on synoptic scales \citep{Vallis2017AtmosphericEdition, Gill1982Atmosphere-oceanDynamics,Pedlosky1987GeophysicalDynamics,Zeytounian1990AsymptoticFlows}. At leading order, the horizontal momentum balance reduces to geostrophy, while the vertical momentum equation reduces to hydrostatic balance. Geostrophy implies that the leading-order horizontal flow is divergence-free. At the next order, pseudo-incompressibility (or anelastic continuity) relates the first-order horizontal flow divergence to the vertical derivative of the vertical velocity. Indeed, proceeding to the next order in the Rossby number and eliminating the first-order fields by algebraic manipulations yields a single closed evolution equation for the quasigeostrophic potential vorticity,
\begin{equation} \label{eq: classical-pv}
\frac{\partial q}{\partial t} + \mathbf{v}_g \cdot \nabla q = Q_q,
\end{equation}
with $q$ the potential vorticity, and $\mathbf{v}_g$ the geostrophic velocity derived diagnostically from a streamfunction (see \citet[Ch.~6]{Pedlosky1987GeophysicalDynamics}; \citet{Vallis2017AtmosphericEdition}; \citet{Muraki1999TheTheory}). In this \emph{PV-centric} formulation, all prognostic information is given by two scalar transport equations; Equation \eqref{eq: classical-pv}, the PV transport equation, and the transport equation for the potential temperature evaluated at the bottom boundary (see, e.g., \citet[Ch.~6]{Pedlosky1987GeophysicalDynamics} or \citet[Ch.~5]{Vallis2017AtmosphericEdition} for more details). The three-dimensional velocity and pressure fields are then reconstructed from $q$ by solving an elliptic inversion for the streamfunction of the horizontal flow, subject to appropriate, generally time-dependent, boundary conditions.

This classical PV-based viewpoint proved to be extremely powerful for conceptual and theoretical studies of baroclinic instability, Rossby-wave propagation, and large-scale eddy dynamics \citep{Pedlosky1987GeophysicalDynamics,Salmon1998LecturesDynamics,Vallis2017AtmosphericEdition}. However, its formulation in terms of these two prognostic scalars and the associated elliptic inversion is rather remote from the variable sets and operator structures used in contemporary compressible or soundproof dynamical cores.

In particular, the standard QG equations are usually expressed in pressure coordinates, often utilizing Boussinesq or shallow-water representations, and cannot be directly translated to the Exner-pressure, density-weighted potential-temperature, and flux-form momentum variables used in many numerical models for the atmosphere, including BK19. Embedding classical QG directly into a compressible multimodel solver would therefore require a substantial change of variables and operators, altering the fundamental structure of the BK19 approach.

\citet{Klein2022QG-DL-Ekman:Framework}, based on an asymptotic framework that was progressively evolved throughout a larger body of work (see, among others, \cite{Botta1999DryAsymptotics,Klein2001DistinguishedFlows, Klein2004An, Klein2005MultipleFlows,Klein2008AnAsymptotics,Klein2010RegimeModels,Baumer2023ScalingAir, Baumer2025PQG-DL-Ekman:Closures, Baumer2025AFlows}) proposed a scale-dependent framework to fully describe the dynamics of the quasi-geostrophic regime. Starting from the full compressible equations written in terms of horizontal and vertical velocities, Exner pressure $\pi$, and perturbation potential temperature $\tilde{\Theta}$, the analysis introduces a common nondimensionalization together with a singular perturbation framework based on a small parameter $\varepsilon$ that represents the (nondimensional) equator-to-pole potential temperature difference $\Delta \Theta \big|^{\text{eq}}_{\text{p}} \coloneqq \Theta_{\text{eq}} - \Theta_{\text{p}}$, maintained by meridional variations of solar irradiation with latitude (cf.\ \citet{Keller2006ApproximateMotions,Klein2000AsymptoticMethods,Klein2004An,Klein2008AnAsymptotics,Frierson2008MidlatitudeModels,Schneider2006TheAtmosphere} and the references therein).
By varying the scaling exponents for space and time, one recovers the soundproof, geostrophic, anelastic, and hydrostatic models as distinguished limits of a single transformed equation set \citep{Klein2010Scale-dependentFlows,Klein2008AnAsymptotics, Klein2004An}.

On synoptic scales, with horizontal length $L = h_{sc} / \varepsilon^2$ and advective time $T = h_{sc} / (\varepsilon^2 u_{\mathrm{ref}})$, the Coriolis and pressure-gradient terms become asymptotically dominant and form a geostrophic balance between the dynamical parts of the variables, while the vertical momentum equation reduces to hydrostatic balance:
\begin{equation}
f_0 \,\mathbf{k} \times \mathbf{v}_{\shortparallel} + \nabla_{\shortparallel} \tilde{\pi} = 0,
\qquad
\partial_z \tilde{\pi} = \tilde{\Theta},
\end{equation}
where $\mathbf{v}_\shortparallel$ denotes the horizontal velocity and $\tilde{\pi}$ the leading-order Exner-pressure perturbation \citep[eqs.\ (14a,b)]{Klein2010Scale-dependentFlows}. At the same order, the continuity equation yields the pseudo-incompressible (anelastic-type) divergence constraint for the momenta. Proceeding to the next order in $\varepsilon$ and eliminating higher-order corrections leads to a transport equation for the leading-order potential vorticity
\begin{equation}
q \;=\; \zeta \;+\; \beta \,\eta \;+\; f\,\rho^{-1} \partial_z \!\left( \frac{\rho \tilde{\Theta}}{\dif \bar{\Theta} / \dif z} \right),
\end{equation}
where $\zeta$ is the vertical component of vorticity and $\beta \eta$ accounts for planetary-vorticity variations \citep[eqs.\ (16)–(19)]{Klein2010Scale-dependentFlows}. In this formulation, the QG model emerges as a \emph{balanced, soundproof, hydrostatic limit} of the compressible system, expressed in the same primitive variables $(\mathbf{v}_\shortparallel, w, \tilde{\pi}, \tilde{\Theta})$ that also underlie the anelastic and weak-temperature-gradient regimes \citep[sections 2.2.1–2.2.2]{Klein2010Scale-dependentFlows}.

From the numerical perspective, this formulation is different from the traditional one in two important ways. First, the quasi-geostrophic regime is within a group of asymptotic models, all derived from the common governing equations; the QG PV equation is not a starting point but a derived consequence of geostrophic and hydrostatic balance under synoptic scaling. Second, the leading-order balances and the prognostic information about the dynamics are expressed directly in terms of the usual variables of a typical dynamical core, e.g., density, momenta, potential temperature, and pressure rather than in terms of a scalar PV (together with the mentioned potential temperature equation) and a streamfunction. This makes the QG limit structurally compatible with soundproof and hydrostatic formulations in Exner pressure and mass-weighted potential temperature \citep{Klein2010Scale-dependentFlows}.

In the present work, we adopt this scale-dependent viewpoint. We use the quasigeostrophic limit identified by \citet{Klein2022QG-DL-Ekman:Framework} as the analytical target for a new set of \emph{geostrophic switches} within BK19. In doing so, we retain the BK19 variable set and semi-implicit structure for time integration while enforcing discrete geostrophic and hydrostatic balances consistent with the QG asymptotics. The associated Exner-pressure elliptic equation \citep[eq.\ (19)]{Klein2010Scale-dependentFlows} emerges at the discrete level as a consequence of the momentum and potential temperature transport equations. This is in contrast to traditional QG codes, which discretize the PV equation directly and solve a separate elliptic problem for the streamfunction.

\subsection{Related numerical schemes and balanced models}

Semi-implicit and asymptotic-preserving schemes for compressible atmospheric dynamics have received considerable attention over the past decades (\citet{Klein2001AsymptoticMechanics, Marras2015AGalerkin, Mengaldo2019CurrentPrediction}). Operational dynamical cores such as the ECMWF IFS, based on the hydrostatic primitive equations (\citet{Temperton2001AModel}; \citet{Hortal2002TheModel}), and the Met Office ENDGame core (\citet{Wood2014AnEquations}; \citet{Benacchio2016Semi-implicitPerspective}; \citet{Melvin2019AGeometry}) employ semi-implicit or semi-Lagrangian methods to overcome the stiffness associated with fast acoustic and internal gravity waves. Other frameworks, such as the EULAG model with its MPDATA-based finite-volume advection (\citet{Prusa2008EULAGFlows}; \citet{Smolarkiewicz2014ADynamics, Smolarkiewicz2016AFlows, Smolarkiewicz2019Semi-implicitDynamics}) and related finite-volume dynamical cores (\citet{Kuhnlein2019FVMIFS}), incorporate soundproof or quasi-hydrostatic variants of the governing equations as submodels but do not aim for a strictly seamless transition between compressible and reduced regimes in the sense of a single blended discretization.

\subsection{Contribution}

By combining the BK19 numerical framework with the scale-dependent asymptotic analysis of \citet{Klein2022QG-DL-Ekman:Framework}, this paper makes the following contributions:

\begin{itemize}
  \item \textbf{Embedding of modern quasi-geostrophic theory into BK19.} We derive a quasi-geostrophic limit of the compressible Euler equations in the BK19 variables by utilizing the asymptotic setup of \citet{Klein2022QG-DL-Ekman:Framework} in the synoptic scale and small Rossby number and directly applying it on the numerical scheme. The resulting system combines geostrophic and hydrostatic balance with pseudo-incompressibility, from which the QG Poisson equation for pressure is recovered as a consequence.

  \item \textbf{Design of geostrophic switches compatible with existing soundproof and hydrostatic switches.} Building on this analytical limit, we introduce new blending parameters and modifications of the BK19 implicit substep that allow access to the quasi-geostrophic balance regime from the main variables and equations while preserving the original structure of the numerical discretization. This results in a discretization that covers several important regimes in atmospheric dynamics, namely, the full compressible, pseudo-incompressible, hydrostatic, and quasi-geostrophic modes within the same finite-volume framework, without changing the basic numerical operators.

  \item \textbf{Consistent discrete curl operator in the BK19 framework.} Using the discrete derivative operators defined in BK19, we construct a discrete curl operator that is consistent with the underlying staggered-grid structure. This allows us to define the physical vorticity as well as a vorticity-like numerical variable and to show that the resulting pressure equation interacts with this variable in the same way as in the continuous QG limit, while both the physical vorticity and the vorticity-like field remain prognostic throughout the time evolution.

\item \textbf{Numerical results across regimes.}  
Through suites of barotropic and baroclinic vortices, we show how the extended BK19 scheme works consistently in fully compressible behaviour to pseudo-incompressible and finally to quasi-geostrophic dynamics. In the barotropic case, the compressible model exhibits vortex collapse in the weak-rotation regime, whereas the QG configurations preserve the balanced structure. In the baroclinic case, the initial state remains stationary with precision with suppressed internal waves in the QG run, while the full compressible run shows only small deviations created by the internal waves. These results highlight that the scheme accurately captures the QG limit and preserves the rate of convergence of the original scheme. This validates that the scheme now can blend into different regimes consistently from within a single discretization framework.

\end{itemize}

The paper is organized as follows. Section~\ref{sec: governing-equations} recalls the governing equations and their discretization following BK19. Section~\ref{sec: time-integration-scheme} presents the time-integration scheme proposed of BK19, including the reformulation of the governing equations (Section~\ref{subsec: reformulation-of-the-governing-equations}), the evolution equations for the perturbation variables (Section~\ref{subsec: evolution-equation-for-perturbation-variables}), the semi-implicit discretization (Section~\ref{subsec: semi-implicit}), and a detailed analysis of the implicit Euler step (Section~\ref{subsec: implicit-step-details}). Section~\ref{sec: access-to-quasi-geostrophic-regime} discusses in detail how to obtain the quasi-geostrophic equations from within the numerical framework and how to incorporate the corresponding switches that provide access to the QG regime. First we introduce the discrete curl operator (Section~\ref{subsec: discrete-curl-operator}), then we carry out the asymptotic scale analysis (Section~\ref{subsec: asymptotic-analysis}), and finally, we introduce the switches in Section~\ref{subsec: introducing-corresponding-switches}. Section~\ref{sec: numerical-results} describes the numerical experiments and analyzes the performance of the extended algorithm in the QG regime and in transitions between regimes. Section~\ref{sec: discussions-and-conclusions} summarizes the results and discusses the conclusions.

\section{Governing equations}
\label{sec: governing-equations}

We consider adiabatic, compressible flow of an ideal gas in a rotating frame of reference, under the $f$-plane approximation and subject to a constant gravitational acceleration $g$. The governing equations can be written in flux form as
\begin{subequations}\label{eq:governing-equations}
\begin{align}
\rho_{t} + \nabla_{\shortparallel}\cdot(\rho\mathbf{u}) + (\rho w)_{z} &= 0,
\label{eq: governing-density}
\\
(\rho\mathbf{u})_{t} + \nabla_{\shortparallel}\cdot(\rho\mathbf{u}\circ\mathbf{u}) + (\rho w\mathbf{u})_{z} &=
- \bigl( c_p P \nabla_{\shortparallel}\pi + f\,\mathbf{k}\times\rho\mathbf{u}\bigr),
\label{eq: governing-horizontal-momentum}
\\
(\rho w)_{t} + \nabla_{\shortparallel}\cdot(\rho\mathbf{u} w) + (\rho w^2)_{z} &=
- \bigl( c_p P\,\pi_{z} + \rho g \bigr),
\label{eq:governing-vertical-momentum}
\\ P_{t} + \nabla_{\shortparallel}\cdot(P\mathbf{u}) + (P w)_{z} &= 0,
\label{eq: governing-pressure}
\end{align}
\end{subequations}
where $\rho$ is the density, $\mathbf{u}$ the horizontal velocity vector, $w$ the vertical velocity component, and $P$ the mass-weighted potential temperature. The Exner pressure $\pi$ and $P$ are related to the thermodynamic pressure $p$ via
\begin{align} \label{eq: EOS}
\pi = \left(\frac{p}{p_{\mathrm{ref}}}\right)^{R/c_p},
\qquad
P = \rho \Theta = \frac{p_{\mathrm{ref}}}{R} \left( \frac{p}{p_{\mathrm{ref}}}\right)^{c_v/c_p},
\end{align}
with $c_p$ and $c_v$ the specific heat capacities at constant pressure and volume, respectively, $R = c_p - c_v$ the gas constant, $p_{\mathrm{ref}}$ a reference pressure, and $\Theta$ the potential temperature. Here $\nabla_{\shortparallel}$ denotes the horizontal gradient operator, the symbol $(\cdot)_z$ the partial derivative with respect to $z$, and $\mathbf{u}\circ\mathbf{u}$ the dyadic product.

And finally, combining \eqref{eq: governing-density}, \eqref{eq: governing-pressure}, and \eqref{eq: EOS} yields the familiar advection equation for the potential temperature:
\begin{align} \label{eq: governing-potential-temperature}
    \Theta_t + \mathbf{u} \cdot \nabla_{\shortparallel} \Theta  + w\,\Theta _z = 0.
\end{align}

\section{Time Integration Scheme}
\label{sec: time-integration-scheme}
This section is devoted to describe BK19 the numerical scheme which joins the ideas developed in \citet{Klein2009AsymptoticsEquations, Benacchio2014ADynamics, Klein2016ADynamics} and borrows ideas from \citet{Smolarkiewicz1993OnFramework, Smolarkiewicz1997OnFlows, Smolarkiewicz2014ADynamics}.

\subsection{Reformulation of the governing Equations}
\label{subsec: reformulation-of-the-governing-equations}

As in the continuous formulation, the prognostic variables in the BK19 model are the density, horizontal and vertical momentum, and mass-weighted potential temperature, i.e.\ $(\rho, \rho\mathbf{u}, \rho w, P)$. By introducing two blending parameters $\alpha_P$ and $\alpha_w$, the scheme provides a seamless transition between the compressible and pseudo-incompressible equations \citep{Durran1989ImprovingApproximation,Klein2010Scale-dependentFlows} and between the non-hydrostatic and hydrostatic limits \citep{Benacchio2014ADynamics,Klein2016ADynamics}.

The authors, following the same path as in \citet{Klein2016ADynamics}, introduce the inverse potential temperature
\begin{align}
    \chi = \frac{1}{\Theta},
\end{align}
and the governing equations are reformulated as
\begin{subequations} \label{eq: BK19-governing-equations}
\begin{align}
\rho_{t} + \nabla_{\shortparallel}\cdot (P \mathbf{u}\,\chi) + (P w \chi)_z &= 0,
\label{eq: BK19-density}
\\
(\rho\mathbf{u})_{t} + \nabla_{\shortparallel}\cdot\bigl(P\mathbf{u} \circ \chi \mathbf{u}\bigr) + (P w \chi \mathbf{u})_z 
&= -\bigl(c_p P\nabla_{\shortparallel}\pi + f\,\mathbf{k} \times \rho \mathbf{u}\bigr),
\label{eq: BK19-horizontal-momentum}
\\
\alpha_w\bigl[(\rho w)_{t} + \nabla_{\shortparallel}\cdot (P\mathbf{u} \chi w) + (P w \chi w)_z\bigr] 
&= -\bigl(c_p P \pi_z + \rho g\bigr),
\label{eq: BK19-vertical-momentum}
\\
\alpha_P P_t + \nabla_{\shortparallel} \cdot (P \mathbf{u}) + (P w)_z &= 0,
\label{eq: BK19-pressure}
\end{align}
\end{subequations}
where the vector field
\[
P\mathbf{v} = (P\mathbf{u}, P w)
\]
acts as the advecting flux in this formulation.

\subsection{Evolution equations for perturbation variables}
\label{subsec: evolution-equation-for-perturbation-variables}

For implicit numerical methods that involve compressibility, Earth rotation, and buoyancy, it is essential to represent the large-scale background stratification of potential temperature separately from the evolving perturbation fields. A common strategy in the literature is therefore to introduce perturbation quantities around a prescribed background state, and the baseline version of the BK19 algorithm, outlined in this section, adopts this viewpoint.

In particular, the main separation between background and perturbation fields is introduced for the Exner pressure and the potential temperature (or, in the present formulation, its inverse). We write
\begin{align}
    \pi(t, \mathbf{x}, z) &= \overline{\pi}(z) + \pi^{\prime}(t, \mathbf{x}, z), \\
    \chi(t, \mathbf{x}, z) &= \overline{\chi}(z) + \chi^{\prime}(t, \mathbf{x}, z),
\end{align}
where an overbar denotes the background profiles and a prime denotes perturbations. The background fields satisfy the hydrostatic balance
\begin{align} \label{eq: hydrostatic-background-equation}
    \frac{\mathrm{d} \overline{\pi}}{\mathrm{d} z} = - \frac{g}{c_p}\, \overline{\chi},
    \qquad 
    \overline{\pi}(0) = 1.
\end{align}

Using the equation of state \eqref{eq: EOS} and the fact that the background Exner pressure is time independent, the evolution equation for $P$, cf.\ \eqref{eq: BK19-pressure}, can be rewritten in terms of the Exner pressure perturbation as
\begin{align}\label{eq: BK19-Exner-pressure}
    \alpha_P \left( \frac{\partial P}{\partial \pi} \right) \pi^{\prime}_t 
    = - \nabla \cdot \bigl[P(\pi)\,\mathbf{v}\bigr].
\end{align}

Similarly, the mass continuity equation \eqref{eq: BK19-density} can be rewritten as an evolution equation for the perturbation of the inverse potential temperature,
\begin{align}\label{eq: BK19-chi-perturbation}
    (P\chi^{\prime})_t 
    + \nabla_{\shortparallel} \cdot (P \mathbf{u}\,\chi^{\prime}) 
    + (P w \chi^{\prime})_z 
    = - P w\, \overline{\chi}_z
\end{align}
which makes explicit the coupling between the perturbation field and the background stratification.

\subsection{Semi-implicit time discretization}
\label{subsec: semi-implicit}

For a more compact formulation, BK19 adopt the notation of \citet{Smolarkiewicz2014ADynamics} and introduce the vector of cell variables
\begin{align} \label{eq: psi-notation}
    \Psi = \bigl(\chi,\, \chi \boldsymbol{u},\, \chi w,\, \chi^{\prime}\bigr),
\end{align}
so that the system \eqref{eq: BK19-governing-equations} can be written in the compact form
\begin{subequations} \label{eq: compact-governing-equation}
\begin{align}
    (P \Psi)_t + \mathcal{A}(\Psi; P\mathbf{v}) &= \mathcal{Q}(\Psi; P), 
    \label{eq: compact-cell-variables-equation} \\
    \alpha_P P_t + \nabla \cdot (P\mathbf{v}) &= 0.
    \label{eq: compact-pressure-equation}
\end{align}
\end{subequations}
Here \eqref{eq: compact-pressure-equation} is equivalent to \eqref{eq: BK19-Exner-pressure}.

The BK19 algorithm is organized as a two-step semi-implicit scheme. In a first (predictor) step, the advective flux $P\mathbf{v}$ is advanced to the half-time level using a first-order advection scheme. In a second (corrector) step, the variables $\Psi$ and $P$ are updated over the full time step using the predicted fluxes and an implicit trapezoidal rule. In the following we describe these two steps in detail.

\subsubsection{Predictor step}

In the predictor step, the advective fluxes $P\mathbf{v}$ are determined at the half-time level, $(P\mathbf{v})^{n+1/2}$, by discretizing \eqref{eq: compact-pressure-equation} as
\begin{align}
    \alpha_P P^{n+1} = \alpha_P P^n - \Delta t\, \tilde \nabla \cdot (P\mathbf{v})^{n+1/2},
\end{align}
where $\tilde \nabla \cdot$ denotes the discrete divergence operator. For $\alpha_P = 1$, this corresponds to an implicit midpoint update of $P$.

To predict $(P\mathbf{v})^{n+1/2}$, equation \eqref{eq: compact-cell-variables-equation} is split into its advective and nonadvective parts. This is the usual strategy to permit large time steps limited only by the advective Courant number. In a two-stage procedure, the advective update is performed first, followed by an implicit update involving only the source terms:
\begin{subequations} \label{eq: predictor-step}
\begin{align}
    (P \Psi)^{\#} &= \mathcal{A}^{\Delta t / 2}_{\text{1st}} \bigl[\Psi^n ; (P\mathbf{v})^n \bigr], 
    \label{eq: predictor-advection}\\
    P^{\#} &= P^n - \frac{\Delta t}{2}\, \nabla \cdot (P\mathbf{v})^n, 
    \label{eq: predictor-pressure-first}\\
    (P \Psi)^{n+1/2} &= ( P \Psi)^{\#} + \frac{\Delta t}{2}\,\mathcal{Q}(\Psi^{n+1/2}; P^{n+1/2}),
    \label{eq: predictor-variable-implicit}\\
    \alpha_P P^{n+1/2} &= \alpha_P P^{n} - \frac{\Delta t}{2}\, \nabla \cdot (P\mathbf{v})^{n+1/2},
    \label{eq: predictor-pressure-second}
\end{align}
\end{subequations}
where $\mathcal{A}_{1st}^{\Delta t / 2}$ denotes a (at least) first-order advective scheme applied over a half time step. The first two equations provide the explicit advective update, while the last two equations represent an implicit Euler discretization of the nonadvective terms. 

Using \eqref{eq: BK19-Exner-pressure}, equation \eqref{eq: predictor-pressure-second} can be recast as an elliptic problem for the Exner pressure perturbation $\pi^{\prime}$.

\subsubsection{Corrector Step}
In this step, following  \citet{Smolarkiewicz1993OnFramework, Smolarkiewicz1997OnFlows, Smolarkiewicz2014ADynamics}, the authors of BK19 propose to use an implicit trapezoidal rule to update the variables using the precomputed flux from the predictor step. The update rule for the variables are as follows:
\begin{subequations} \label{eq: corrector-step}
        \begin{align}
        &(P \Psi)^{*} = (P \Psi)^{n} + \frac{\Delta t}{2} \mathcal{Q}(\Psi^{n};P^{n}) \label{eq: corrector-explicit} \\
        & (P\Psi)^{**} = \mathcal{A}^{\Delta_t}_{\text{2nd}}[\Psi^{*};(P\mathbf{v})^{n+1/2}] \label{eq: corrector-advection}\\
        &(P \Psi)^{n+1} = (P \Psi)^{**} + \frac{\Delta t}{2} \mathcal{Q}(\Psi^{n+1};P^{n+1}) \label{eq: corrector-implicit}\\
        &\alpha_p (P^{n+1} - P^{n}) = - \Delta t \nabla \cdot (P\mathbf{v})^{n+1/2} \label{eq: corrector-pressure}
    \end{align}
\end{subequations}
Note that equation \eqref{eq: corrector-pressure} is an implicit midpoint update which could have been written as a half-time explicit Euler followed by a half-time implicit Euler updates.

\subsection{Implicit Step Details}
\label{subsec: implicit-step-details}

The reformulations \eqref{eq: psi-notation} and \eqref{eq: compact-governing-equation} allow us to avoid involving $\rho$ and $P$ directly in the updates \eqref{eq: corrector-explicit}–\eqref{eq: corrector-implicit}. Consequently, we only need to evolve $(\mathbf{u}, w, \chi^{\prime})$. The corresponding linearized system for these variables can be written as
\begin{subequations}\label{eq: continuous-implicit-euler-step}
\begin{align}
U_t &= -c_p (P\Theta)^{\circ} \,\pi_x + f V, \\
V_t &= -c_p (P\Theta)^{\circ} \,\pi_y - f U, \\
\alpha_w W_t &= -c_p (P\Theta)^{\circ} \,\pi_z - g\,\frac{\tilde{\chi}}{\chi^{\circ}}, \\
\tilde{\chi}_t &= -W \,\frac{d\bar\chi}{dz}, \\
\alpha_P \left(\frac{\partial P}{\partial\pi}\right)^{\circ} \pi_t &= -\left(U_x + V_y + W_z\right),
\end{align}
\end{subequations}
where $(U, V, W, \tilde \chi) = (P u, P v, P w, P \chi^{\prime})$ and $(P\Theta)^{\circ}$, $\chi^{\circ}$, and $\left(\frac{\partial P}{\partial\pi}\right)^{\circ}$ are precomputed and available at the beginning of the implicit Euler step.

These equations are discretized in time using the precomputed coefficients as
\begin{subequations}\label{eq: discrete-implicit-euler-step}
\begin{align}
U^{n+1} &= U^{**} - \Delta t\, c_p (P\Theta)^{\circ} \,\pi^{n+1}_x + \Delta t\, f\, V^{n+1}, \label{eq: discrete-implicit-momentum-u} \\
V^{n+1} &= V^{**} - \Delta t\, c_p (P\Theta)^{\circ} \,\pi^{n+1}_y - \Delta t\, f\, U^{n+1},\label{eq: discrete-implicit-momentum-v}  \\
\alpha_w W^{n+1} &= \alpha_w W^{**} - \Delta t\, c_p (P\Theta)^{\circ} \,\pi^{n+1}_z - \Delta t\, g\,\frac{\tilde{\chi}^{n+1}}{\chi^{\circ}}, \label{eq: discrete-implicit-momentum-w} \\
\tilde{\chi}^{n+1} &= \tilde{\chi}^{**} - \Delta t\, \frac{d \bar\chi}{dz}\, W^{n+1},\label{eq: discrete-implicit-chi}  \\
\alpha_P \left(\frac{\partial P}{\partial\pi}\right)^{\circ}\pi^{n+1} &= \alpha_P \left(\frac{\partial P}{\partial\pi}\right)^{\circ}\pi^{n} -\Delta t \left(U^{n+1}_x + V^{n+1}_y + W^{n+1}_z \right),\label{eq: discrete-implicit-pressure} 
\end{align}
\end{subequations}
where $(\cdot)^{**}$ denotes the explicitly updated state. One should note that the discretization separates the variables into cell-centered quantities $(U, V, W, \tilde \chi, P)$ and the node-based variable $\pi$ (see Section~\ref{subsec: discrete-curl-operator} for details).

Solving \eqref{eq: discrete-implicit-momentum-u}–\eqref{eq: discrete-implicit-chi} for the variables at time level $n+1$, we obtain
\begin{subequations} \label{eq: full-momenta-equations}
\begin{align} 
    \begin{pmatrix}
        U \\
        V
    \end{pmatrix}^{n+1} 
    &= \frac{1}{1 + \left(\Delta t f\right)^2}\left[
    \begin{pmatrix}
         U + \left(\Delta t f\right) V \\
         V - \left(\Delta t f\right) U
    \end{pmatrix}^{**}
     - \Delta t c_p (P\Theta)^{**}
    \begin{pmatrix}
         \pi_x + \left(\Delta t f\right) \pi_y \\
         \pi_y - \left(\Delta t f\right) \pi_x
    \end{pmatrix}^{n+1}\right], 
    \label{eq: discrete-solved-momentum-horizontal}\\
    W^{n+1} &= \frac{1}{\alpha_w + \left(\Delta t N\right)^2} \left[ \alpha_w W^{**} - \Delta t g \frac{\tilde \chi^{**}}{\chi^{\circ}} - \Delta t c_p (P\Theta)^{**} \pi_z^{n+1} \right],
    \label{eq: discrete-solved-momentum-w}
\end{align}
\end{subequations}
where
\begin{equation}
N^2 = -g\,\frac{1}{\chi^{\circ}}\,\frac{d \bar \chi}{dz}.
\end{equation}

Substituting \eqref{eq: discrete-solved-momentum-horizontal}–\eqref{eq: discrete-solved-momentum-w} into the pressure equation \eqref{eq: discrete-implicit-pressure} yields a Helmholtz-type equation for $\pi^{n+1}$:
{\small
\begin{align}\label{eq: helmholtz-pressure-system}
    &\alpha_P\left(\frac{\partial P}{\partial \pi}\right)^{\circ}  \pi^{n+1} 
    - \Delta t^2 \Bigg\{
        \left[ \frac{c_p(P\Theta)^{\circ}}{1 + (\Delta t f)^2}  \bigl( \pi_x^{n+1} + \Delta t f\, \pi_y^{n+1}\bigr)\right]_x 
        + \left[\frac{c_p(P\Theta)^{\circ}}{1 + (\Delta t f)^2} \bigl( \pi_y^{n+1} - \Delta t f\, \pi_x^{n+1}\bigr)\right]_y \notag + \left[ \frac{c_p(P\Theta)^{\circ}}{\alpha_w + (\Delta t N)^2} \pi_z^{n+1} \right]_z\Bigg\}\\
    &=  \alpha_P\left(\frac{\partial P}{\partial \pi}\right)^{\circ}  \pi^{n} 
    - \Delta t \Bigg\{
        \left[\frac{U^{**} + \Delta t f\, V^{**}}{1 + (\Delta t f)^2} \right]_x 
     + \left[\frac{V^{**} - \Delta t f\, U^{**}}{1 + (\Delta t f)^2} \right]_y
    + \left[\frac{\alpha_w W^{**} - \Delta t g \,\, (\tilde \chi^{**}/ \chi^{\circ})}{\alpha_w + (\Delta t N)^2} \right]_z\Bigg\}.
\end{align}}

\begin{rem}
    The BK19 algorithm and the definitions of the discrete operators are restricted to  Cartesian grids, and the formulation of the equations is not valid for non-Cartesian grids.
\end{rem}

\section{Access to quasi-geostrophic Regime}
\label{sec: access-to-quasi-geostrophic-regime}

In theory, the pseudo-incompressibility switch $\alpha_P$ and the hydrostatic balance switch $\alpha_w$ are obtained from an asymptotic analysis of the governing equations in the limits of small Mach and Froude numbers, respectively. The parameter $\alpha_P$ is associated with a regime in which acoustic waves are asymptotically filtered (small Mach number), while $\alpha_w$ corresponds to a regime where vertical accelerations are negligible in comparison with the vertical pressure gradient and gravity (small Froude number, i.e.\ hydrostatic balance). In both cases, the non-dimensional parameters permit a direct identification of asymptotically small terms in the governing equations, and the resulting reduced systems can be written in closed form for a single set of perturbation variables.

In contrast, access to the quasi-geostrophic regime is not as straightforward. In the pseudo-incompressible and hydrostatic limits, the reduced systems' equations involve only a single asymptotic order of the perturbation variables. By comparison, the QG equations arise from a combination of different orders of the variables (cf., for example, \citet{Klein2022QG-DL-Ekman:Framework} and \citet{Klein2010Scale-dependentFlows}). Specifically, this combination appears in three equations: first, the diagnostic geostrophic balance equation in which the Coriolis force from the leading-order velocity field balances the pressure gradient in the leading dynamical order (typically second-order in the perturbation expansions); second, in the potential temperature evolution equation; and finally, in the vertical vorticity equation in which different asymptotic orders of the horizontal velocity are combined. As a result, the QG limit is not obtained by a simple elimination of small terms but rather by a systematic separation of orders across the asymptotic equation hierarchy. Consequently, one cannot expect a single, direct QG switch at the level of the governing equations that would be analogous to $\alpha_P$ and $\alpha_w$.

In this section, we study the model's behavior in the QG regime by applying the usual asymptotic scale analysis of the continuous model directly to the numerical algorithm. Instead of first deriving a continuous QG model and discretizing it, the discretized BK19 scheme is the starting point, and the asymptotic scale analysis is applied directly to it. This is the expected behavior of asymptotic-preserving discretizations: the discrete scheme should reduce, in the limit regime, to a consistent discretization of the QG equations without changing the underlying discretization and operators.

This derivation consists of three steps. First, we introduce a discrete curl operator that is consistent with the discrete gradient operator used in the BK19 model and that allows us to define  the vorticity in its numerical and physical forms. The numerical vorticity variable will serve as the counterpart of the vertical component of physical vorticity in the BK19 numerical scheme, which is central to QG dynamics. Second, the asymptotic scale analysis of \citet{Klein2022QG-DL-Ekman:Framework} is applied to the BK19 discretized equations to derive the QG Poisson equation for the Exner pressure. In this way, the elliptic problem solved in the semi-implicit step is reduced to the QG inversion problem under the QG scaling. Finally, two additional switches are introduced into \eqref{eq: discrete-solved-momentum-horizontal} in order (i) to obtain  the corresponding elliptic problem and (ii) to control the update of the horizontal momenta. These switches enable a (seamless) blending between the full compressible dynamics and its QG limit within a single numerical framework.

\subsection{Discrete curl operator}
\label{subsec: discrete-curl-operator}

\begin{figure}
    \centering
    \includegraphics[width=0.5\linewidth]{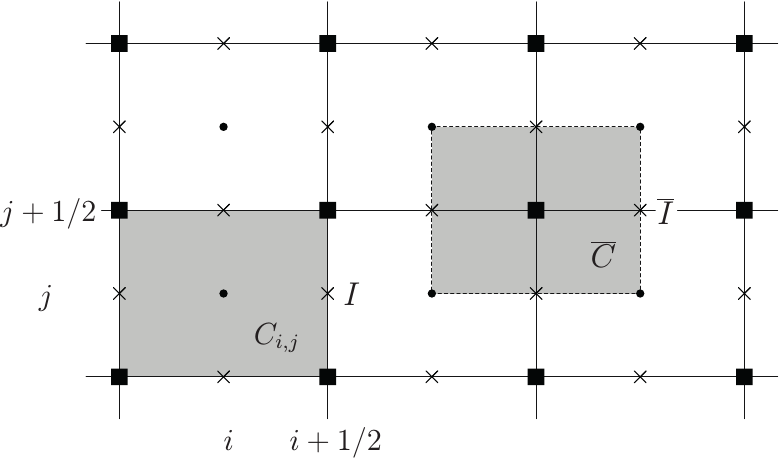}
    \caption{Grid arrangement for a two dimensional slice. $C_{i,j,k}$: the primary control volumes, $\bullet$: the primary cell centers, $\blacksquare$: the dual grid centers (nodes), $\times$: the cell/node interfaces.}
    \label{fig: grid-arrangement}
\end{figure}

The BK19 algorithm defines the primary variables $(U, V, W, \tilde\chi, P)$ at the centers of the control volumes $C_{i, j, k}$ of a Cartesian grid with constant grid spacing and indices $i = 0, \dots, I-1$, $j = 0, \dots, J-1$, and $k = 0, \dots, K-1$ in the three coordinate directions. A two-dimensional slice of this grid is shown in Fig.~\ref{fig: grid-arrangement}. The control volumes $C_{i,j,k}$ form the primary grid, and the corresponding cell centers (denoted by $\bullet$ in the figure) are the locations at which the prognostic variables are evaluated. The dual grid (nodes, denoted by $\blacksquare$) centers are where the pressure variable is evaluated.

In this setting, the discrete derivative in $x$ of a cell-centered variable $U$, evaluated at the grid nodes, is given by \citet[eq.~31]{Benacchio2019ADynamics} as
\begin{subequations}
\label{eq: discrete-derivative-on-cells}
\begin{align}
\left(U_x\right)_{i+\frac{1}{2},\, j+\frac{1}{2},\, k+\frac{1}{2}}
&= \frac{1}{\Delta x}
   \left(
       \hat{U}_{i+1,\, j+\frac{1}{2},\, k+\frac{1}{2}}
       - \hat{U}_{i,\, j+\frac{1}{2},\, k+\frac{1}{2}}
   \right), \\[0.5em]
\hat{U}_{i,\, j+\frac{1}{2},\, k+\frac{1}{2}}
&= \frac{1}{4}\Big(
       U_{i,\, j+1,\, k+1}
       + U_{i,\, j,\, k+1}
       + U_{i,\, j+1,\, k}
       + U_{i,\, j,\, k}
   \Big).
\end{align}
\end{subequations}
Here $\hat{U}_{i,\, j+\frac{1}{2},\, k+\frac{1}{2}}$ denotes the average of the four cell-centered values surrounding the point $(i,\, j+\tfrac{1}{2},\, k+\tfrac{1}{2})$, which lies on an interface of the dual grid. The derivative $\left(U_x\right)_{i+\frac{1}{2},\, j+\frac{1}{2},\, k+\frac{1}{2}}$ is then approximated by a centered finite difference of these edge-averaged values. This construction can be viewed as a two-stage procedure: interpolation from cell centers to interfaces, followed by a centered difference in the $x$-direction. The derivatives in the other spatial directions are defined analogously using the corresponding edge averages in $y$ or $z$.

Similarly, the derivative in $x$ of a node-based variable $\pi^{\prime}$, evaluated at cell centers, is given by \citet[eq.~30]{Benacchio2019ADynamics} as
\begin{subequations}
\label{eq: discrete-derivative-on-nodes}
\begin{align}
\left(\pi^{\prime}_x\right)_{i,\, j,\, k}
&= \frac{1}{\Delta x}
   \left(
       \hat{\pi}^{\prime}_{i+\frac{1}{2},\, j,\, k}
       - \hat{\pi}^{\prime}_{i-\frac{1}{2},\, j,\, k}
   \right), \\[0.5em]
\hat{\pi}^{\prime}_{i+\frac{1}{2},\, j,\, k}
&= \frac{1}{4}\Big(
       \pi^{\prime}_{i+\frac{1}{2},\, j+\frac{1}{2},\, k+\frac{1}{2}}
       + \pi^{\prime}_{i+\frac{1}{2},\, j-\frac{1}{2},\, k+\frac{1}{2}}
       + \pi^{\prime}_{i+\frac{1}{2},\, j+\frac{1}{2},\, k-\frac{1}{2}}
       + \pi^{\prime}_{i+\frac{1}{2},\, j-\frac{1}{2},\, k-\frac{1}{2}}
   \Big).
\end{align}
\end{subequations}
In this case, $\pi^\prime$ is given at the nodes of the grid, and \eqref{eq: discrete-derivative-on-nodes} provides a centered approximation of its derivative at the primary cell centers. As in \eqref{eq: discrete-derivative-on-cells}, the operator first computes a local average and then applies a centered difference. 

We now introduce a discrete curl operator compatible with this framework. The primary requirement is that this operator reproduces, at the discrete level, the continuous identity
\[
\mathbf{k}\cdot(\nabla\times\mathbf{u}) = v_x - u_y,
\]
while remaining consistent with the discrete derivatives defined above. In addition, the curl is to be defined at the nodes, where the Exner pressure $\pi$ resides, so that we can achieve $\nabla \times \nabla \pi = 0$. Here we only focus on the vertical component of the curl (for which we have a direct use case in the equations); the remaining components can be defined analogously.

Let $\mathbf{u} = (U, V, W)$ denote a three-dimensional field. 
The vertical component of the curl is defined at the grid nodes by
\begin{align}
 \textbf{curl}_k (\mathbf{u})_{i+\frac{1}{2},\, j+\frac{1}{2},\, k+\frac{1}{2}}
 &:= \big[ \mathbf{k} \cdot (\nabla \times \mathbf{u}) \big]_{i+\frac{1}{2},\, j+\frac{1}{2},\, k+\frac{1}{2}} \notag\\
 &\phantom{:}= \frac{1}{2} \Big(
        (\mathbf{k} \cdot \nabla \times \mathbf{u})_{i+\frac{1}{2},\, j+\frac{1}{2},\, k}
        + (\mathbf{k} \cdot \nabla \times \mathbf{u})_{i+\frac{1}{2},\, j+\frac{1}{2},\, k+1}
    \Big),
\end{align}
where
\begin{align}
    \big[\mathbf{k} \cdot \nabla \times \mathbf{u}\big]_{i+\frac{1}{2},\, j+\frac{1}{2},\, k}
    &= \frac{1}{\Delta x} \big( \tilde V_{i+1,\, j+\frac{1}{2},\, k} - \tilde V_{i,\, j+\frac{1}{2},\, k} \big)
      - \frac{1}{\Delta y} \big( \tilde U_{i+\frac{1}{2},\, j+1,\, k} - \tilde U_{i+\frac{1}{2},\, j,\, k} \big),
\end{align}
and the discrete components $\tilde U$ and $\tilde V$ are given by
\begin{align}
    \tilde U_{i+\frac{1}{2},\, j,\, k} &= \frac{1}{2}\big( U_{i+1,\, j,\, k} + U_{i,\, j,\, k} \big), \\
    \tilde V_{i,\, j+\frac{1}{2},\, k} &= \frac{1}{2}\big( V_{i,\, j+1,\, k} + V_{i,\, j,\, k} \big).
\end{align}
 The expression for $[\mathbf{k} \cdot \nabla \times \mathbf{u}]_{i+\frac{1}{2},\, j+\frac{1}{2},\, k}$ therefore approximates $v_x - u_y$ at vertical level $k$, while the subsequent average over $k$ and $k+1$ centers the curl at the node $(i+\tfrac{1}{2},\, j+\tfrac{1}{2},\, k+\tfrac{1}{2})$ in the vertical direction.

In \ref{app: curl-properties} it is shown that 
\begin{align}\label{eq: discrete-curl-raw}
\textbf{curl}_k (\mathbf{u})_{i+\frac{1}{2},\, j+\frac{1}{2},\, k+\frac{1}{2}} = (V_x)_{i+\frac{1}{2},\, j+\frac{1}{2},\, k+\frac{1}{2}} - (U_y)_{i+\frac{1}{2},\, j+\frac{1}{2},\, k+\frac{1}{2}}.
\end{align}
That is, when the definitions of the discrete derivatives $(V_x)$ and $(U_y)$ from \eqref{eq: discrete-derivative-on-cells} are used, the above construction reduces exactly to the difference of these discrete derivatives. This identity demonstrates that the discrete curl operator is consistent with the existing discrete gradient operator in the BK19 scheme, in the sense that it reproduces at the discrete level the same algebraic structure as the continuous identity $\mathbf{k}\cdot\nabla\times\mathbf{u} = v_x - u_y$.

\begin{rem}
If $\mathbf{u}$ denotes the discrete velocity field, then \eqref{eq: discrete-curl-raw} has the form of the vertical component of vorticity evaluated at the grid nodes. In the present numerical scheme, however, the prognostic variables are $(U, V, W, \tilde{\chi}) = (P u, P v, P w, P \chi^{\prime})$, so that \eqref{eq: discrete-curl-raw} represents a vorticity-like \emph{numerical} quantity rather than the physical vorticity itself. In particular, because $U = P u$ and $V = P v$, the combination $(V_x - U_y)$ involves horizontal derivatives of the weighted momentum components, not of the velocity components alone.

In asymptotic analysis (see section \ref{subsec: asymptotic-analysis}), in the leading order, where $P_0(z) = \rho_0(z)$, one may identify
\[
Z_0 \coloneqq V_{0,x} - U_{0,y} = \rho_0\,\zeta_0 ,
\]
so that $Z_0$ (the leading-order vertical component of the \emph{numerical} vorticity variable) corresponds to the mass-weighted vertical component of the \emph{physical} vorticity, denoted by $\zeta_0$. This observation is central to deriving a prognostic equation for the physical vorticity within this framework (see Section~\ref{sec: prognostic-evolution-of-vorticity} for details).

In general, however, the discrete curl applied on the numerical momentum should be interpreted as a \emph{pseudo-vorticity} associated with the staggered, weighted representation of momentum. As demonstrated in subsequent sections, this pseudo-vorticity exhibits behaviour closely analogous to the continuous vertical vorticity in the limit of small Rossby number and therefore plays an important role in accessing the quasi-geostrophic regime at the discrete level.
\end{rem}

\subsection{Asymptotic analysis of the discretized equations}
\label{subsec: asymptotic-analysis}

\subsubsection{Quasi-geostrophic scaling and nondimensionalization of the update equations}

Access to the QG regime is obtained by analysing the limit of the discrete update equations for small Rossby number. To do this in a systematic manner, the asymptotic scale analysis must be incorporated explicitly into the numerical algorithm rather than being applied only at the level of the continuous equations. The modern theory of quasi-geostrophic dynamics offers a variety of such asymptotic expansions and scaling analyses; see, for example, \citet{Klein2022QG-DL-Ekman:Framework, Klein2010Scale-dependentFlows, Klein2000AsymptoticMethods, Arakawa2009UnificationEquations, Klein2008AnAsymptotics, Botta1999DryAsymptotics}. In the present work we adopt the asymptotic setup introduced in \citet{Klein2022QG-DL-Ekman:Framework} for the discrete setting.

For completeness, we recall the non-dimensionalization used therein. Dimensional quantities are denoted by a superscript asterisk:
\begin{equation}
\pi = \left( \frac{p^*}{p_{\mathrm{ref}}} \right)^{\kappa},
\qquad
\rho = \frac{\rho^*}{\rho_{\mathrm{ref}}},
\qquad
\Theta = \frac{\Theta^*}{\Theta_{\mathrm{ref}}},
\qquad
\mathbf{u} = \frac{\mathbf{u}^*}{u_{\mathrm{ref}}},
\qquad
w = \frac{w^*}{u_{\mathrm{ref}}}\,\frac{\ell_{\mathrm{ref}}}{h_{\mathrm{sc}}}.
\end{equation}
Here $p_{\mathrm{ref}}$, $\rho_{\mathrm{ref}}$ and $\Theta_{\mathrm{ref}}$ are suitably chosen reference values, $u_{\mathrm{ref}}$ is a characteristic horizontal velocity scale, and $\ell_{\mathrm{ref}}$ and $h_{\mathrm{sc}}$ denote characteristic horizontal and vertical length scales, respectively. The vertical velocity is scaled anisotropically, reflecting the small aspect ratio typical of large-scale atmospheric flows.

The corresponding nondimensional control parameters are the Mach, external Froude, and Rossby numbers,
\begin{equation}
\mathrm{M} = \frac{u_{\mathrm{ref}}}{\sqrt{p_{\mathrm{ref}}/\rho_{\mathrm{ref}}}},
\qquad
\mathrm{Fr}_{\mathrm{ext}} = \frac{u_{\mathrm{ref}}}{\sqrt{g h_{\mathrm{sc}}}},
\qquad
\mathrm{Ro} = \frac{u_{\mathrm{ref}}}{f_0 \,\ell_{\mathrm{ref}}},
\end{equation}
and we nondimensionalize space and time according to
\begin{equation}
t = \frac{t^{*} u_{\mathrm{ref}}}{\ell_{\mathrm{ref}}}, 
\qquad
\boldsymbol{x} = (x, y) = \frac{\boldsymbol{x}^{*}}{\ell_{\mathrm{ref}}}, 
\qquad
z = \frac{z^{*}}{h_{\mathrm{sc}}}.
\end{equation}

Following \citet{Klein2022QG-DL-Ekman:Framework}, we adopt the distinguished limits
\begin{equation}
\mathrm{M}^{2} = \mathrm{Fr}^{2}_{\mathrm{ext}} = \varepsilon^{3},
\qquad
\mathrm{Ro} = \frac{\varepsilon}{f_{0}},
\qquad
\frac{h_{\mathrm{sc}}}{\ell_{\mathrm{ref}}} = \varepsilon^{2},
\end{equation}
which together with $N_{\mathrm{ref}}^2 \; h_{\mathrm{sc}}/g = \varepsilon$ characterize the QG scaling regime of interest. Under these distinguished limits, the leading-order dynamics are quasi-geostrophic, with compressibility and vertical acceleration effects entering at higher orders.

Adopting the scaling assumptions given above, the system \eqref{eq: discrete-implicit-euler-step} can be cast in nondimensional form as
\begin{subequations}\label{eq:NDDiscreteImplicitEuler}
\begin{align}
U^{n+1} &= U^{**} - \frac{1}{\varepsilon^3}\,\Delta t\, c_p (P\Theta)^{\circ} \,\pi^{n+1}_x 
         + \frac{1}{\varepsilon}\,\Delta t\, f\, V^{n+1}, \\
V^{n+1} &= V^{**} - \frac{1}{\varepsilon^3}\,\Delta t\, c_p (P\Theta)^{\circ} \,\pi^{n+1}_y 
         - \frac{1}{\varepsilon}\,\Delta t\, f\, U^{n+1}, \\
W^{n+1} &= W^{**} - \frac{1}{\varepsilon^7}\,\Delta t\, c_p (P\Theta)^{\circ} \,\pi^{n+1}_z 
         - \frac{1}{\varepsilon^7}\,\Delta t\, g\,\frac{\tilde{\chi}^{n+1}}{\chi^{\circ}}, \\
\tilde{\chi}^{n+1} &= \tilde{\chi}^{**} + \varepsilon\,\Delta t\, \frac{\chi^{\circ}}{g} N^2\, W^{n+1}, \\
\alpha_P \left(\frac{\partial P}{\partial\pi}\right)^{\circ} \pi^{n+1} 
&= \alpha_P \left(\frac{\partial P}{\partial\pi}\right)^{\circ} \pi^{n} 
   - \Delta t \left(U^{n+1}_x + V^{n+1}_y + W^{n+1}_z \right).
\end{align}
\end{subequations}
Here, the superscript $n+1$ denotes the implicitly updated time level, $U^{**}$, $V^{**}$, $W^{**}$, and $\tilde{\chi}^{**}$ are the contributions from the explicit step, and $(P\Theta)^{\circ}$, $\chi^{\circ}$ denote the values available at the beginning of the implicit step. 

Solving the momentum equations in \eqref{eq:NDDiscreteImplicitEuler} for $U^{n+1}$, $V^{n+1}$, and $W^{n+1}$ yields
\begin{subequations}\label{eq: momenta-solved}
\begin{align}
\bigl[\varepsilon^2 + (\Delta t\, f)^2\bigr] U^{n+1} 
&= \varepsilon^2 U^{**} 
   - \frac{1}{\varepsilon}\,\Delta t\, c_p (P\Theta)^{\circ} \,\pi^{n+1}_x 
   - \frac{1}{\varepsilon^2}\,\Delta t^2 f\, c_p (P\Theta)^{\circ} \,\pi^{n+1}_y 
   + \varepsilon\, \Delta t\, f\, V^{**}, \\
\bigl[\varepsilon^2 + (\Delta t\, f)^2\bigr] V^{n+1} 
&= \varepsilon^2 V^{**} 
   - \frac{1}{\varepsilon}\,\Delta t\, c_p (P\Theta)^{\circ} \,\pi^{n+1}_y 
   + \frac{1}{\varepsilon^2}\,\Delta t^2 f\, c_p (P\Theta)^{\circ} \,\pi^{n+1}_x 
   - \varepsilon\, \Delta t\, f\, U^{**}, \\
\bigl[\varepsilon^6 + (\Delta t\, N)^2\bigr] W^{n+1} 
&= \varepsilon^6 W^{**} 
   - \frac{1}{\varepsilon}\,\Delta t\, c_p (P\Theta)^{\circ} \,\pi^{n+1}_z 
   - \frac{1}{\varepsilon}\,\Delta t\, g\,\frac{\tilde{\chi}^{**}}{\chi^{\circ}}. \label{eq: momenta-solved-z}
\end{align}
\end{subequations}
These relations explicitly show the balance between the implicit forcing due to pressure gradients and buoyancy and the contributions of $(\Delta t\,f)^2$ and $(\Delta t\,N)^2$.

For the asymptotic analysis we follow \citet{Klein2022QG-DL-Ekman:Framework} to introduce the following QG expansions of the variables:
\begin{subequations}\label{eq: QG-expansions}
\begin{align}
\rho(t, \mathbf{x}, z) 
&= \rho_0(z) + \varepsilon\,\rho_1(z) + \varepsilon^2 \rho^{(2)}(t, \mathbf{x}, z) + \mathcal{O}(\varepsilon^3), \\
\pi(t, \mathbf{x}, z) 
&= \pi_0(z) + \varepsilon\,\pi_1(z) + \varepsilon^2 \pi^{(2)}(t, \mathbf{x}, z) + \mathcal{O}(\varepsilon^3), \\
\Theta(t, \mathbf{x}, z) 
&= 1 + \varepsilon\,\Theta_1(z) + \varepsilon^2 \Theta^{(2)}(t, \mathbf{x}, z) + \mathcal{O}(\varepsilon^3),\label{eq: asymptotic-expansion-Theta} \\
\mathbf{u}(t, \mathbf{x}, z) 
&= \mathbf{u}^{(0)}(t, \mathbf{x}, z) + \varepsilon\,\mathbf{u}^{(1)}(t, \mathbf{x}, z) + \mathcal{O}(\varepsilon^{2}), \\
w(t, \mathbf{x}, z) 
&= \varepsilon\, w^{(1)}(t, \mathbf{x}, z) + \mathcal{O}(\varepsilon^2),
\end{align}
\end{subequations}
where $\mathbf{u}$ denotes the horizontal velocity. The expansions separate background hydrostatic profiles, encoded in the $z$-dependent coefficients $(\rho_0,\pi_0,\Theta_0)$ and their $\mathcal{O}(\varepsilon)$ corrections, from the $\mathcal{O}(\varepsilon^2)$ perturbations that host the QG dynamics. The velocity expansions reflect the standard QG assumption that the horizontal velocity has a non-trivial leading-order component $\mathbf{u}^{(0)}$ (geostrophic flow), whereas the vertical velocity is of smaller magnitude, of order $\varepsilon$.

Inspired by these expansion ansatzes, and keeping the expansions for $\pi$ and $\Theta$ unchanged, we adopt the following quasi-geostrophic expansions for the numerical variables $(U, V, W, \tilde \chi)$ and $P$:
\begin{subequations}\label{eq: QG-numerical-expansion}
    \begin{align}
        U(t, \mathbf{x}, z) &= U_{0}(t, \mathbf{x}, z) + \varepsilon\, U_{1}(t, \mathbf{x}, z) + \mathcal{O}(\varepsilon^2), \\
        V(t, \mathbf{x}, z) &= V_{0}(t, \mathbf{x}, z) + \varepsilon\, V_{1}(t, \mathbf{x}, z) + \mathcal{O}(\varepsilon^2), \\
        W(t, \mathbf{x}, z) &= \varepsilon\, W_{1}(t, \mathbf{x}, z) + \mathcal{O}(\varepsilon^2), \\
        \tilde \chi(t, \mathbf{x}, z) &= 1 + \varepsilon\, \tilde \chi_1(z) + \varepsilon^2\, \tilde \chi^{(2)}(t, \mathbf{x}, z) + \mathcal{O}(\varepsilon^3), \\
        P(t, \mathbf{x}, z) 
&= P_0(z) + \varepsilon\,P_1(z) + \varepsilon^2 P^{(2)}(t, \mathbf{x}, z) + \mathcal{O}(\varepsilon^3), 
    \end{align}
\end{subequations}
where the subscripts $0$ and $1$ refer to the leading-order and first-order contributions in $\varepsilon$, respectively.

\begin{rem}
    The numerical variables in \eqref{eq: QG-numerical-expansion} can be related to the corresponding physical variables in \eqref{eq: QG-expansions} by their definitions, e.g.\ $U = P u = \rho \Theta u$, and similarly for $V$, $W$, $\tilde \chi$ and $P$, followed by expanding each physical quantity in its corresponding powers of $\varepsilon$. In this way, the discrete QG expansions are tightly linked to the underlying continuous asymptotic structure.
\end{rem}

\begin{rem} \label{rem: background-theta-is-one}
    Equation \eqref{eq: asymptotic-expansion-Theta} implies $\Theta_0 \equiv 1$. This implies $P_0 = \rho_0$, since $P_0 = \rho_0 \Theta_0$.
\end{rem}

\subsubsection{Asymptotic balances: numerical vs.\ physical variables}
\begin{rem}
In the sequel we omit the superscript $\circ$ on the precomputed coefficient fields $(P\Theta)^{\circ}$ and $\chi^{\circ}$ for notational brevity; the meaning is unchanged.
\end{rem}

At order $\mathcal{O}(\varepsilon^{-1})$ in equation \eqref{eq: momenta-solved-z}, we obtain
\begin{align} \label{eq: zero-order-numerical-hydrostatic}
    c_p (P\Theta)^{(0)}(\pi_z^{n+1})^{(0)} = -g \frac{(\tilde \chi^{**})^{(0)}}{\chi^{(0)}},
\end{align}
which is equivalent to the following equation in terms of physical variables:
\begin{align} \label{eq: zero-order-physical-hydrostatic}
    c_p (\pi_z^{(0)})^{n+1} = -g (\chi^{\prime^{(0)}})^{**}.
\end{align}
We stress the distinction: \eqref{eq: zero-order-numerical-hydrostatic} is a relation for the leading-order of the numerical field $\tilde \chi$, whereas \eqref{eq: zero-order-physical-hydrostatic} is the same identity expressed in physical variable $\chi^{\prime}$ after substitution and expansion. Since $(\tilde \chi^{(0)})^{n+1} = (\tilde \chi^{(0)})^{**}$, these equation correspond to the hydrostatic balance in the background.
    
At order $\mathcal{O}(1)$ the discrete system recovers the geostrophic and hydrostatic relations for the numerical variables:
\begin{subequations}\label{eq: first-order-balance}
\begin{align}
U^{n+1}_0 &= -\frac{1}{f}\,c_p (P\Theta)^{(0)} \bigl(\pi_y^{(2)}\bigr)^{n+1}, \\
V^{n+1}_0 &= \phantom{-}\frac{1}{f}\,c_p (P\Theta)^{(0)} \bigl(\pi_x^{(2)}\bigr)^{n+1}, \\
\bigl(\frac{\tilde{\chi}^{**}}{\chi}\bigr)^{(1)} &= -\frac{1}{g}c_p \bigl[(P\Theta) \pi_{z}^{n+1}\bigr]^{(1)}. \label{eq: first-order-numerical-hydrostatic}
\end{align}
\end{subequations}
The first two equations express discrete geostrophic balance between the Coriolis force and horizontal pressure gradients at order $\varepsilon^0$, whereas the third relation represents the hydrostatic balance.

Equation \eqref{eq: first-order-numerical-hydrostatic} governs the numerical variables at leading order. Expanding the terms and utilizing \eqref{eq: zero-order-physical-hydrostatic} yields the corresponding leading-order equation for the physical variables:
\begin{align} \label{eq: first-order-physical-hydrostatic}
        c_p (\pi_z^{(1)})^{n+1} = -g (\chi^{\prime^{(1)}})^{**}
\end{align}
again noting that $(\tilde \chi^{(1)})^{n+1} = (\tilde \chi^{(1)})^{**}$, these equations are the hydrostatic balance in the background deviations.

Proceeding to order $\mathcal{O}(\varepsilon)$ gives the ageostrophic corrections for the numerical variables:
\begin{subequations}\label{eq: second-order-balance}
\begin{align}
U^{n+1}_1 &= -\frac{1}{\Delta t\, f^2} c_p (P\Theta)^{(0)}\bigl(\pi_x^{(2)}\bigr)^{n+1}
            - \frac{1}{f} c_p (P\Theta)^{(0)}\bigl(\pi_y^{(3)}\bigr)^{n+1}  - \frac{1}{f} c_p (P\Theta)^{(1)}\bigl(\pi_y^{(2)}\bigr)^{n+1}
            + \frac{1}{\Delta t\, f} V^{**}_0, \label{eq: second-order-balance-u}\\
V^{n+1}_1 &= -\frac{1}{\Delta t\, f^2} c_p (P\Theta)^{(0)}\bigl(\pi_y^{(2)}\bigr)^{n+1}
            + \frac{1}{f} c_p (P\Theta)^{(0)}\bigl(\pi_x^{(3)}\bigr)^{n+1} + \frac{1}{f} c_p (P\Theta)^{(1)}\bigl(\pi_x^{(2)}\bigr)^{n+1}
            - \frac{1}{\Delta t\, f} U^{**}_0,  \label{eq: second-order-balance-v}\\
W^{n+1}_1 &= - \frac{1}{\Delta t\, N^2} c_p \bigl[(P\Theta) \pi_z^{n+1}\bigr]^{(2)}
            - \frac{g}{\Delta t\, N^2}\,\bigl[\frac{\tilde{\chi}^{**}}{\chi} \bigr]^{(2)}. \label{eq: second-order-balance-w}
\end{align}
\end{subequations}

The vertical relation \eqref{eq: second-order-balance-w} can be mapped back into physical variables to yield the first physically meaningful hydrostatic balance at this order,
 \begin{align}
 \label{eq: second-order-physical-hydrostatic}
        c_p (\pi_z^{(2)})^{n+1} = -g (\chi^{\prime^{(2)}})^{n+1}.
 \end{align}

\begin{rem}
    Equations \eqref{eq: zero-order-numerical-hydrostatic} and \eqref{eq: first-order-numerical-hydrostatic} are the hydrostatic balance in the non-dynamical background state and therefore are not informative. Equations \eqref{eq: second-order-balance-w} is the first equation for hydrostatic balance containing the dynamical parts of the variables.
\end{rem}

\subsubsection{Divergence constraint and QG inversion.}
To derive the discrete Helmholtz equation associated with the QG regime, we first consider the divergence of the discrete momenta:
\begin{align}
\label{eq: divergence-constraint-expanded}
U^{n+1}_x + V^{n+1}_y + W^{n+1}_z 
&= \bigl(U^{n+1}_{0, x} + V^{n+1}_{0, y} + W^{n+1}_{0, z}\bigr) + \varepsilon \bigl(U^{n+1}_{1, x} + V^{n+1}_{1, y} + W^{n+1}_{1, z}\bigr)
   + \mathcal{O}(\varepsilon^{2}).
\end{align}
The leading-order contribution vanishes because $(U_0, V_0)$ is geostrophic (and thus horizontally divergence-free) and $W_0 \equiv 0$ in the expansion \eqref{eq: QG-expansions}. Moreover, under the synoptic QG scaling the coefficient fields $(P\Theta)^{(0)}$ and $(P\Theta)^{(1)}$ depend only on $z$, hence horizontal mixed-derivative terms stemming from their expansion cancel. Collecting the $\mathcal{O}(\varepsilon)$ terms and substituting \eqref{eq: second-order-balance} gives the numerical divergence relation
\begin{align}
U^{n+1}_{1, x} + V^{n+1}_{1, y} + W^{n+1}_{1, z} 
&= - \frac{1}{\Delta t\, f^2} \Bigl\{ c_p\Bigl[(P\Theta)(\pi_x^{n+1})\Bigr]^{(2)}_x 
+ c_p\bigl[(P\Theta)(\pi_y^{n+1})\Bigr]^{(2)}_y\Bigr\} \notag\\
&\quad - \frac{1}{\Delta t}\, c_p\Bigl[\frac{(P\Theta)}{N^2}\,(\pi_z^{n+1})\Bigr]^{(2)}_z 
       + \frac{1}{\Delta t\, f}\Bigl(V^{**}_{0, x} - U^{**}_{0, y}\Bigr)
       - \frac{g}{\Delta t}\,\Bigl[\frac{\tilde{\chi}^{**}}{N^2\chi} \Bigr]^{(2)}_z.
\end{align}
This expression relates the divergence of the ageostrophic velocity to horizontal and vertical pressure-gradient contributions, the vorticity and the vertical derivative of the buoyancy perturbation. The first two terms on the right-hand side represent the structure of an anisotropic elliptic operator acting on $\pi^{(2)}$, while the remaining terms act as sources associated with vorticity and stratification.

Restricting to the pseudo-incompressible limit ($\alpha_P=0$) and enforcing zero divergence at $\mathcal{O}(\varepsilon)$ yields following elliptic problem in the numerical variables,
\begin{align} \label{eq: QG-inversion-equation}
c_p \Bigl[(P\Theta)(\pi_x^{n+1})\Bigr]^{(2)}_x 
+ c_p \Bigl[(P\Theta)(\pi_y^{n+1})\Bigr]^{(2)}_y
+ c_p f^2\Big[\frac{(P\Theta)}{N^2}\,(\pi_z^{n+1}) \Bigr]^{(2)}_z
= f\,Z^{**}_0 - g f^2\,\Bigl[\frac{\tilde{\chi}^{**}}{N^2\chi} \Bigr]^{(2)}_z,
\end{align}
where we introduced the numerical pseudo-vorticity
\begin{align} \label{eq: vorticity-numerical}
    Z^{**}_0 \coloneqq V^{**}_{0,x} - U^{**}_{0,y} = P_0\,(v_{0,x}-u_{0,y}).
\end{align}

Equation \eqref{eq: QG-inversion-equation} is the discrete Poisson-type equation for the Exner-pressure perturbation in the QG limit; its left-hand side defines an anisotropic elliptic operator consistent with the QG elliptic operator, while the right-hand side couples pressure to (numerical) vorticity and stratification anomalies.

\begin{rem}
    Equation \eqref{eq: QG-inversion-equation} is formulated in terms of numerical variables; this expansion sufficiently clarifies the corresponding QG switches within the numerical algorithm (see Section \ref{subsec: introducing-corresponding-switches}). Physically, however, the terms can be expanded to verify equivalence with the QG Poisson equation for the physical variables:
    \begin{align}
        \label{eq: QG-inversion-equation-physical}
        c_p \nabla_{\shortparallel}^{2} (\pi^{(2)})^{n+1}
+ \frac{c_p \; f^2}{\rho_0}\Big[ \frac{\rho_0}{ N^2}(\pi_z^{(2)})^{n+1} \Bigr]_z
= f\,\zeta^{**}_0 - \frac{g f^2}{\rho_0}\,\Bigl[\frac{\rho_0(\chi^{\prime(2)})^{**}}{N^2} \Bigr]_z,
    \end{align}
    where $\zeta_0 = v_{0, x} - u_{0, y}$ denotes the leading-order physical vertical vorticity.

\begin{rem}
    The bottom boundary condition for the equations \eqref{eq: QG-inversion-equation} and \eqref{eq: QG-inversion-equation-physical} is obtained by determining $d \pi / dz$ directly through the corresponding vertical momentum equation, namely equation \eqref{eq: second-order-balance-w}. In general, the vertical boundary condition of the elliptic operator in the BK19 model is obtained by determining $d \pi / d z$ using equations \eqref{eq: discrete-implicit-momentum-w} and \eqref{eq: discrete-implicit-chi}.
\end{rem}

\end{rem}

\subsubsection{Prognostic evolution of leading-order vorticity.}
\label{sec: prognostic-evolution-of-vorticity}
Finally, we examine the asymptotic evolution of the leading-order vertical pseudo-vorticity $Z_0$. Taking the vertical component of the curl of the leading-order horizontal momentum equations \eqref{eq: first-order-balance} and employing the horizontal divergence of \eqref{eq: second-order-balance-u} and \eqref{eq: second-order-balance-v}, we obtain
\begin{align}
Z_0^{n+1} = Z_0^{**} - \Delta t f\, \nabla_{\shortparallel} \cdot \begin{pmatrix}
    U_1 \\ V_1
\end{pmatrix}^{n+1}.
\end{align}
Since $Z_0 = P_0\zeta_0$, we arrive at the corresponding evolution equation for the physical vorticity:
\begin{align}
    \label{eq: vorticity-evolution-physical}
    \zeta_0^{n+1} = \zeta_0^{n} - \Delta t f \, \nabla_{\shortparallel} \cdot \mathbf{u}^{(1)}.
\end{align}
This demonstrates that the leading-order vorticity remains a prognostic quantity, with its discrete evolution driven by the divergence of the first-order ageostrophic flow. In the continuous limit, this update relation is consistent with the QG vorticity equation, where vorticity (specifically, its vertical component) evolves under the influence of ageostrophic fluxes (see, e.g., \citet[eq.~58]{Klein2008AnAsymptotics} or \citet[eq.~31a]{Klein2022QG-DL-Ekman:Framework}).

\subsection{Introducing corresponding switches in the full equations}
\label{subsec: introducing-corresponding-switches}

The asymptotic analysis in Section~\ref{subsec: asymptotic-analysis} shows that, under QG scaling, the discrete scheme approaches the QG regime, where the horizontal momenta are diagnosed from the geostrophic balance update, and the pressure is obtained by solving a Poisson equation. In contrast, away from the QG regime (e.g., for moderate Rossby numbers or strongly ageostrophic flow), the full compressible or pseudo-incompressible dynamics with prognostic momentum equations must be retained. The next step is therefore to introduce a mechanism in the fully dimensional momentum equations \eqref{eq: full-momenta-equations} that allows one to switch smoothly between these regimes at the discrete level.

In contrast to the incompressible and hydrostatic limits, this is not straightforward in the QG regime, because different orders of variables are combined to create the QG Poisson equation and to make vorticity prognostic. Thus, a direct ``QG switch'' on the momentum equations analogous to the pseudo-incompressibility or hydrostatic switches would not be meaningful. Instead, one must distinguish the behavior of the momentum equation in creating the elliptic problem for the pressure variable from its behavior in updating the horizontal momentum.

To bring these two distinct behaviors into the momentum equation, we introduce two independent switches:
\begin{itemize}
    \item $\alpha_H \in \{0,1\}$, which controls the type of elliptic problem for the Exner pressure (turning it into the QG Poisson problem in the appropriate limit);
    \item $\alpha_u \in \{0,1\}$, which controls whether the implicit update of the horizontal momentum is active (prognostic) or effectively replaced by a diagnostic balance relation.
\end{itemize}

With these switches, the horizontal and vertical momentum updates in equation \eqref{eq: full-momenta-equations} can be written in compact form as
\begin{subequations}\label{eq: full-switches}
\begin{align}
\begin{pmatrix}
U \\
V
\end{pmatrix}^{n+1}
&=
\frac{1}{\alpha_H \alpha_u + \left(\Delta t f\right)^2}
\left[
\alpha_u
\begin{pmatrix}
\alpha_H U^{**} + \left(\Delta t f\right) V^{**} \\
\alpha_H V^{**} - \left(\Delta t f\right) U^{**}
\end{pmatrix}
-
\Delta t c_p (P\Theta)^{\circ}
\begin{pmatrix}
\alpha_u \,\pi_x^{n+1} + \left(\Delta t f\right) \pi_y^{n+1} \\
\alpha_u \,\pi_y^{n+1} - \left(\Delta t f\right) \pi_x^{n+1}
\end{pmatrix}
\right],
\\[0.5em]
W^{n+1}
&=
\frac{1}{\alpha_w + \left(\Delta t N\right)^2}
\left[
\alpha_w W^{**}
- \Delta t\, g\, \frac{\tilde{\chi}^{**}}{\chi^{\circ}}
- \Delta t\, c_p (P\Theta)^{\circ}\, \pi_z^{n+1}
\right],
\end{align}
\end{subequations}
where $(\cdot)^{**}$ as before denotes the state after the explicit step, and $\alpha_w$ is the vertical hydrostatic switch in the previous subsection.

The parameter $\alpha_H$ influences the horizontal update through the explicit updates $U^{**}$ and $V^{**}$, and implicitly through the elliptic equation for $\pi^{n+1}$. The elliptic equation, as shown in the previous section, is obtained by substituting \eqref{eq: full-switches} into the pseudo-compressibility equation. For $\alpha_H = 1$, the explicit part of the momentum, in the form of its divergence, appears in the elliptic operator, and the resulting equation is of the full Helmholtz type, as in the original BK19 scheme. For $\alpha_H = 0$, these contributions are removed from the elliptic operator, and what remains is the vorticity from the explicit part of the momenta as in the QG theory and the corresponding elliptic problem reduces to the Poisson problem derived in Section~\ref{subsec: asymptotic-analysis}.

The parameter $\alpha_u$ directly controls whether the horizontal momentum is updated prognostically or diagnostically using the geostrophic balance equation. For $\alpha_u = 1$, the explicitly updated fields $U^{**}$ and $V^{**}$ enter the right-hand side of \eqref{eq: full-switches} and are corrected by the implicit pressure-gradient and Coriolis terms, yielding a fully prognostic semi-implicit update. For $\alpha_u = 0$, the terms involving $U^{**}$ and $V^{**}$ are dropped completely, resulting in the geostrophic balance equation to determine $(U,V)^{n+1}$ which is consistent with the leading-order relations \eqref{eq: first-order-balance}.

The vertical update remains independent of the new switches and is controlled by the hydrostatic switch in BK19 through $\alpha_w$. When $\alpha_w = 1$, the full semi-implicit vertical momentum equation is used; when $\alpha_w = 0$, the vertical velocity is diagnosed from the hydrostatic balance between $\pi_z$ and $\tilde{\chi}$, as in the asymptotic limit discussed in Section~\ref{subsec: asymptotic-analysis}. On the other hand, the pseudo-incompressible $\alpha_P$ switch \textit{must} always be set to zero when the new QG switches are to be used to guarantee that the elliptic equation is a Poisson type equation and not a Helmholtz type equation.
 
To activate the solver in the QG regime, the switches are applied in two logical stages. First, we enforce the pseudo-incompressible and hydrostatic limits by setting $\alpha_P = 0$ and $\alpha_w = 0$, and choosing $\alpha_H = 0$, $\alpha_u = 1$ in \eqref{eq: full-switches}. The resulting Poisson equation is solved for the pressure correction $\pi^{n+1}$. In a second stage, after $\pi^{n+1}$ is determined, we restore $\alpha_H = 1$, set $\alpha_u = 0$, and reuse \eqref{eq: full-switches} to compute $(U,V)^{n+1}$.

In summary, choosing $\alpha_H = \alpha_u = 1$ recovers the original fully implicit momentum update with a Helmholtz elliptic problem, while other combinations of $(\alpha_H,\alpha_u)$ allow one to select independently whether to construct the QG-type elliptic problem for the pressure and whether to use the resulting pressure field to update the horizontal momentum. It is important to note that, in order to activate the QG regime, both switches should not be set to zero simultaneously, as doing so would \textit{not} yield the QG elliptic equation. We emphasize once again that the QG switches introduced above are always tied to the soundproof switch $\alpha_P$ in the sense that a QG flow is only admissible when $\alpha_P = 0$, ensuring that pseudo-incompressibility is guaranteed and the Helmholtz operator is transformed into the Poisson operator. The admissible combinations of $(\alpha_H,\alpha_u,\alpha_w, \alpha_P)$ thus define a family of regimes ranging from fully compressible or pseudo-incompressible dynamics to hydrostatic and QG limits within a single numerical framework.

\section{Numerical Results}
\label{sec: numerical-results}

The switch mechanisms introduced in the previous section were evaluated using two idealized test cases: a barotropic vortex without gravity and a baroclinic vortex. The tests aim to verify the functionality of the switches and compare the QG simulation with the full compressible one. The model is implemented in Python and employs a BiCGSTAB linear solver \citep{VanderVorst1992BI-CGSTAB:Systems}. The computational cost of the QG simulation is  higher than that of the full compressible case, with the baroclinic vortex case (where $g$ is active and the buoyancy and vertical structure of the vortex come into play) considerably higher than that of the barotropic vortex case. The code is available under \url{https://github.com/amir-maghoul/AtmPy}.

\subsection{Barotropic vortex}

For the first test case, we chose a smooth, traveling vortex with non-constant density, which was originally proposed by \citet{Kadioglu2008ADynamics}. This configuration defines a two-dimensional vortex initialized from the gradient-wind balance in the absence of Coriolis forcing. The velocity and density profiles are given by:
\begin{subequations}
\begin{align}
    \rho(x,y,t) &= 
    \begin{cases}
        \frac{1}{2} + \frac{1}{2}(1 - r^{2})^{6} & r < 1, \\
        \frac{1}{2} & \text{otherwise},
    \end{cases} \\
    u(x,y,t) &= 
    \begin{cases}
        -\sin(\theta)(1 - r)^{6} r^{6} + u_{c} & r < 1, \\
        0 & \text{otherwise},
    \end{cases} \\
    v(x,y,t) &= 
    \begin{cases}
        \cos(\theta)(1 - r)^{6} r^{6} + v_{c} & r < 1, \\
        0 & \text{otherwise}.
    \end{cases}
\end{align}
\end{subequations}

Unlike \citet{Kadioglu2008ADynamics}, we do not derive pressure from the gradient-wind relationship. Instead, the pressure field is initialized via the geostrophic balance equation:
\begin{align}
    p(x,y,t) = 
    \begin{cases}
        p(r) - p(1) & r < 1, \\
        0 & \text{otherwise},
    \end{cases}
\end{align}
where the radial pressure profile $p(r)$ is defined by the expansion:
\begin{equation}
\begin{split}
    p(r) =\;& \frac{r^{7}}{7} -\frac{3 r^{8}}{4} + \frac{4 r^{9}}{3} -\frac{r^{10}}{5} -\frac{45 r^{11}}{22} + \frac{3 r^{12}}{4} + \frac{9 r^{13}}{2} -\frac{36 r^{14}}{7} -\frac{11 r^{15}}{5} \\
    &+ \frac{55 r^{16}}{8} -\frac{33 r^{17}}{17} - 4 r^{18} + \frac{58 r^{19}}{19} + \frac{3 r^{20}}{5} -\frac{10 r^{21}}{7} + \frac{4 r^{22}}{11} \\
    &+ \frac{9 r^{23}}{46} -\frac{r^{24}}{8} + \frac{r^{25}}{50}.
\end{split}
\end{equation}

The vortex is initialized in analytical geostrophic balance at the center $\mathbf{x}_{c} = (0,0,0)$ and is advected by a uniform background flow $\mathbf{u}_{c} = (100, 100, 0)\ \si{\meter\per\second}$. We extend the experiment to a three-dimensional barotropic configuration by assuming vertical uniformity. The Coriolis parameter is set to $f = \SI{1e-4}{\second^{-1}}$ with gravity disabled ($g = 0$). The computational domain is defined as
\[
    \Omega = [-0.5, 0.5]^{3}\ \si{\kilo\meter},
\]
with periodic boundary conditions in the horizontal directions and solid walls in the vertical. The grid resolution is $80 \times 80 \times 3$, and the simulation is integrated to $t = \SI{100}{\second}$.

For a weak Coriolis force and the absence of initial gradient-wind balance, the fully compressible model fails to maintain the vortex structure. As a result, we expect the vortex to collapse in the compressible run, while it is preserved in the QG run. 

\begin{figure}[htbp]
    \centering
    \includegraphics[width=0.67\linewidth]{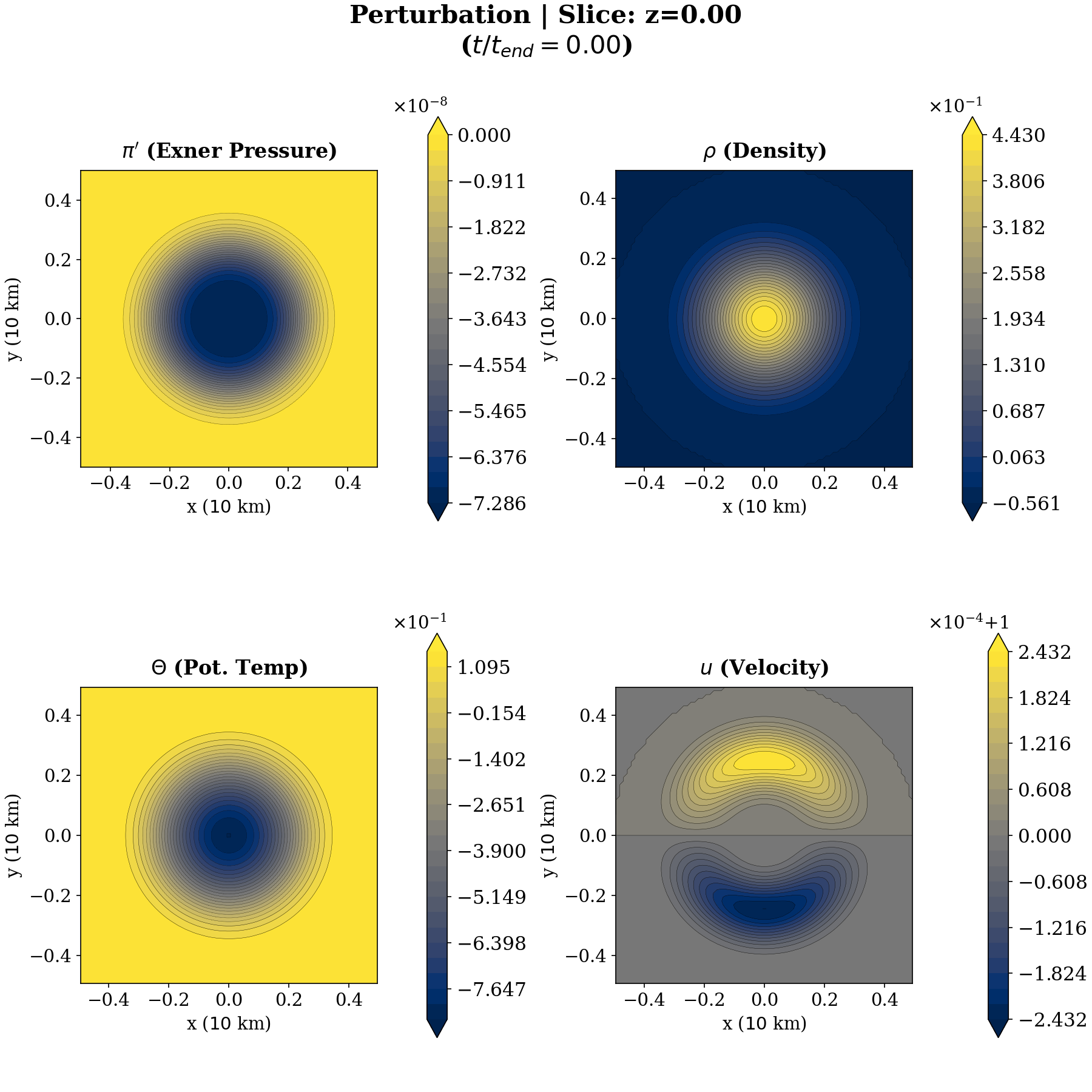}
    \caption{Initial balanced barotropic vortex at $z = 0$: Exner pressure perturbation $\pi'$, density $\rho$, potential temperature $\Theta$, and horizontal velocity $u$.}
    \label{fig:barotropic-initial}
\end{figure}

\begin{figure}[htbp]
    \centering
    \begin{subfigure}[b]{0.31\textwidth}
        \centering
        \includegraphics[width=\linewidth]{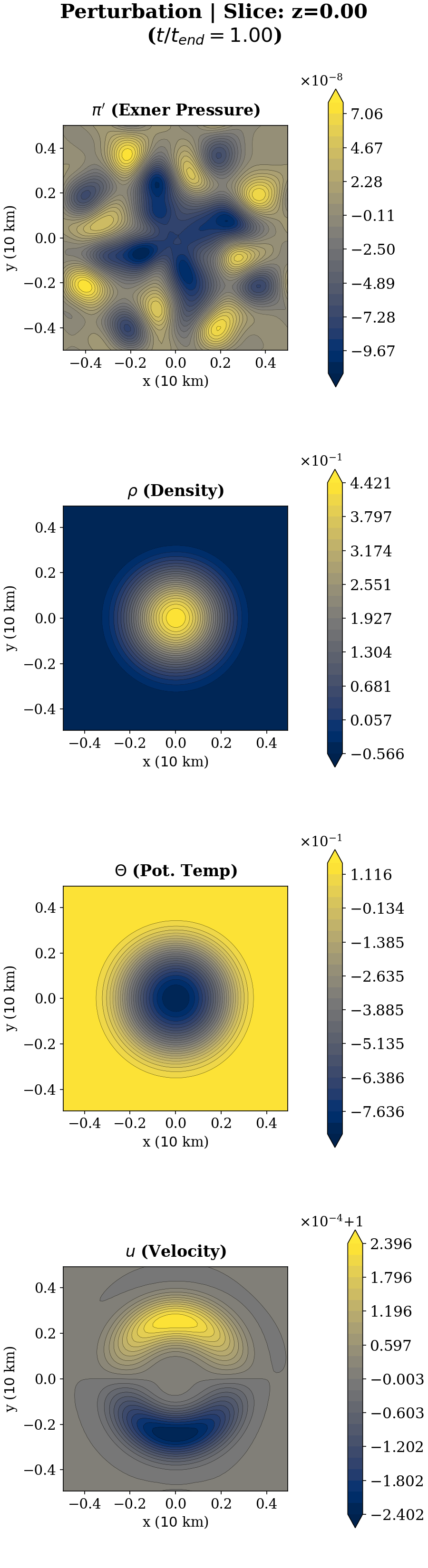}
        \caption{Full compressible}
        \label{fig:end-full}
    \end{subfigure}
    \hfill
    \begin{subfigure}[b]{0.31\textwidth}
        \centering
        \includegraphics[width=\linewidth]{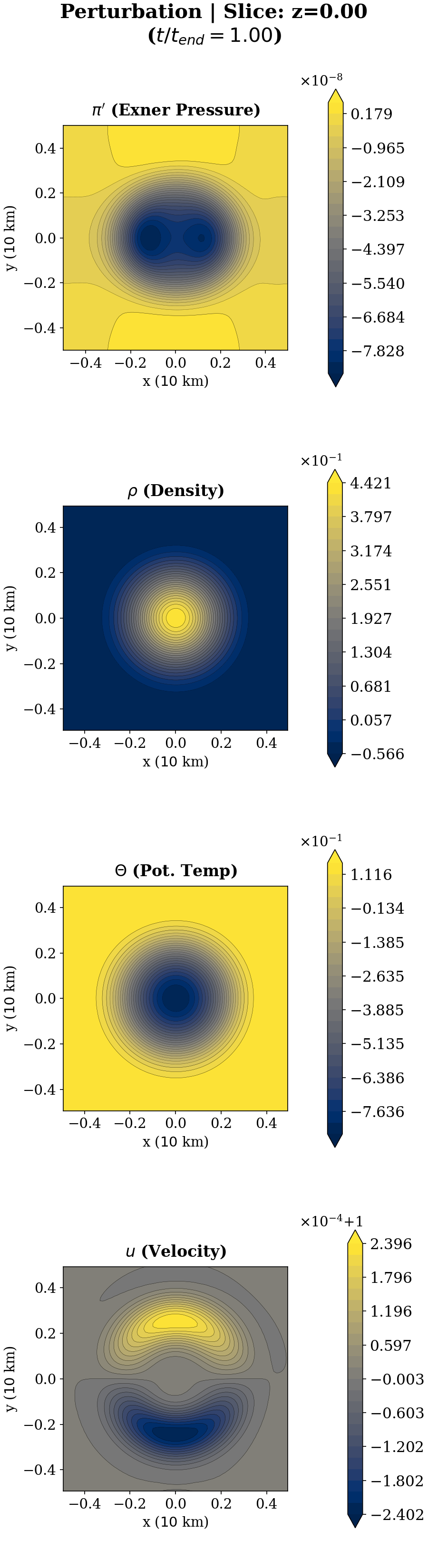}
        \caption{Pseudo-incompressible}
        \label{fig:end-incomp}
    \end{subfigure}
    \hfill
    \begin{subfigure}[b]{0.31\textwidth}
        \centering
        \includegraphics[width=\linewidth]{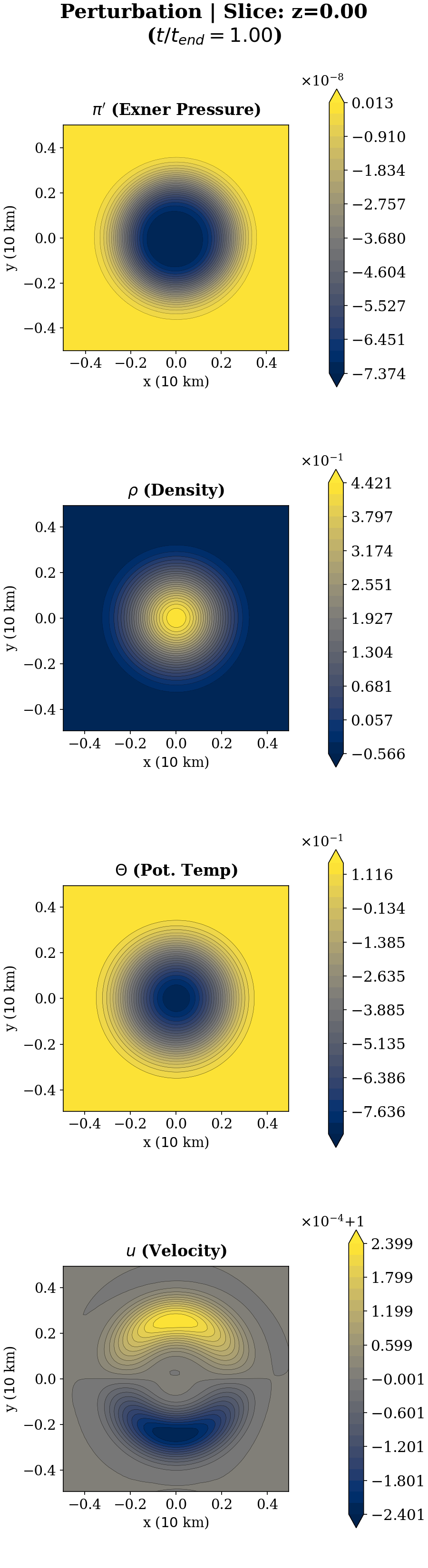}
        \caption{Quasi-geostrophic}
        \label{fig:end-qg}
    \end{subfigure}
    
    \caption{Comparison of vortex states at $t = \SI{100}{\second}$ (after one revolution) for the three different model configurations.}
    \label{fig:barotropic-end-states}
\end{figure}

\begin{figure}[htbp]
    \centering
    \begin{subfigure}[b]{0.31\textwidth}
        \centering
        \includegraphics[width=\linewidth]{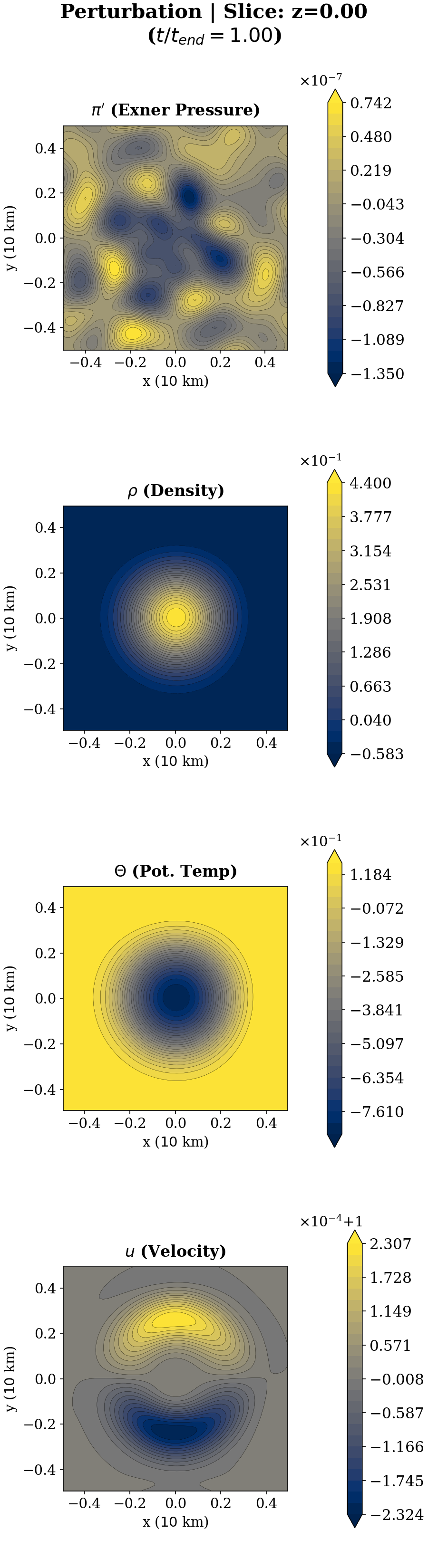}
        \caption{Full compressible}
        \label{fig:end-full_long}
    \end{subfigure}
    \hfill
    \begin{subfigure}[b]{0.31\textwidth}
        \centering
        \includegraphics[width=\linewidth]{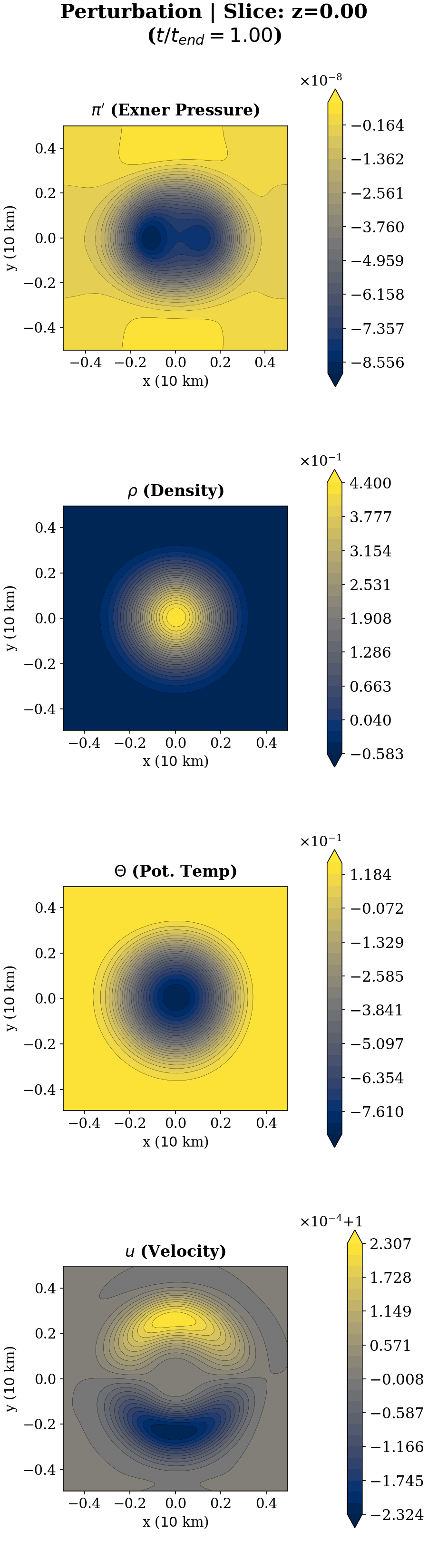}
        \caption{Pseudo-incompressible}
        \label{fig:end-incomp_long}
    \end{subfigure}
    \hfill
    \begin{subfigure}[b]{0.31\textwidth}
        \centering
        \includegraphics[width=\linewidth]{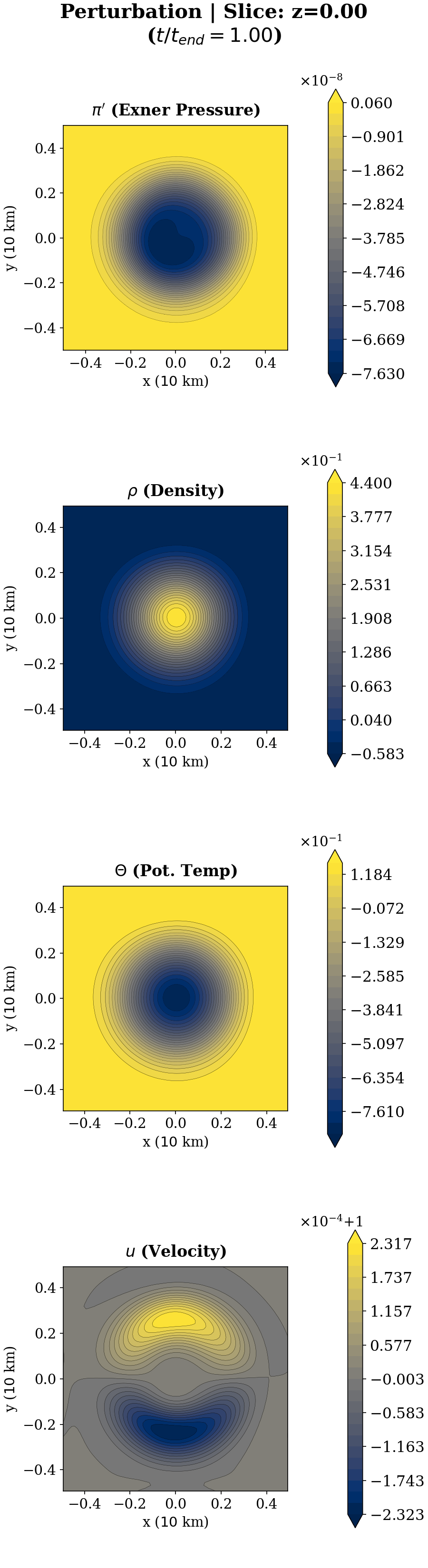}
        \caption{Quasi-geostrophic}
        \label{fig:end-qg_long}
    \end{subfigure}
    
    \caption{Comparison of vortex states at $t = \SI{300}{\second}$ (after three revolutions) for the three different model configurations.}
    \label{fig:barotropic-end-states_long}
\end{figure}

Figure~\ref{fig:barotropic-initial} depicts the initial vortex structure at $z=0$. The simulation results after one horizontal revolution ($t=\SI{100}{\second}$) are compared in Figure~\ref{fig:barotropic-end-states}. As predicted, the fully compressible solution (Fig.~\ref{fig:end-full}) loses structure in the Exner pressure perturbation profile, while the QG model (Fig.~\ref{fig:end-qg}) maintains the vortex structure almost completely. The difference between pseudo-incompressible and QG cases is smaller, since the acoustic waves are filtered in the former case. Figure~\ref{fig:barotropic-end-states_long} depicts the results of a longer simulation, i.e., $t = \SI{300}{\second}$, which is equivalent to three vortex rotations. Here, we can see the start of the collapse of the vortex structure in the pseudo-incompressible case as well (Fig.~\ref{fig:end-incomp_long}), whereas the vortex structure in the QG run still remains intact (Fig.~\ref{fig:end-qg_long}). One might notice the jump in the value of variables in the QG run in comparison to the initial state. This change is a result of the first step geostrophic adjustment of the analytically computed geostrophic balance in the initial conditions to the numerical ones.

\begin{figure}[htbp]
    \centering
    \includegraphics[width=\linewidth]{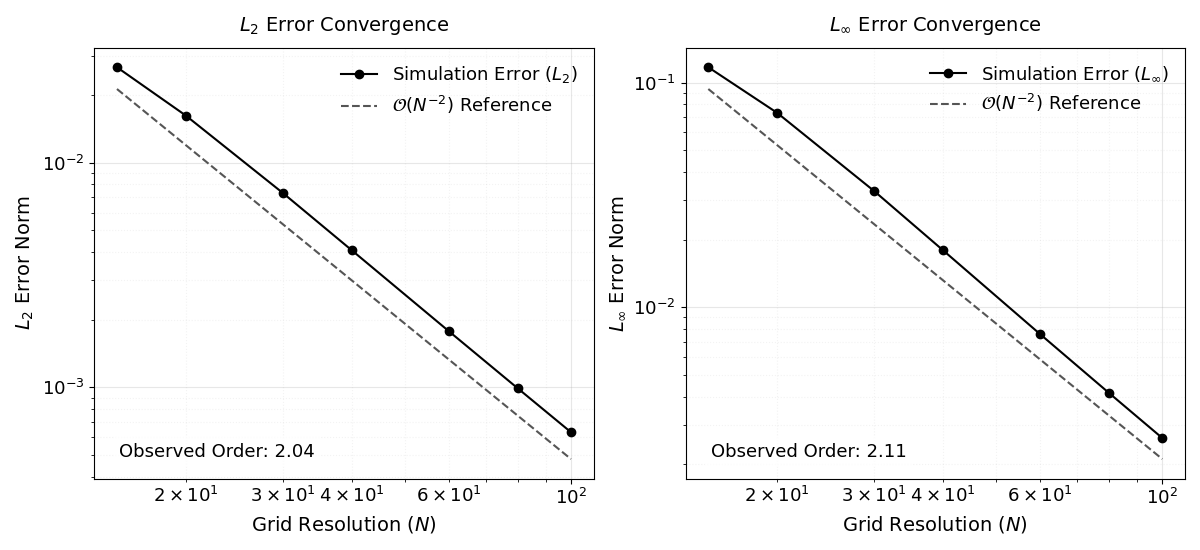}
    \caption{Rate of convergence for the Exner pressure variable. Grid resolution from $15\times 15 \times 15$ to $100 \times 100 \times 100$. (Left) $L_2$-norm and (right) $L_{\infty}$-norm The dashed line displays second-order convergence.}
    \label{fig: convergence}
\end{figure}

Finally, we test the rate of convergence of the scheme in QG mode, to confirm that the switches do not change the expected rate of convergence of the original scheme. Figure~\ref{fig: convergence} shows the $L_2$ and $L_{\inf}$ errors between the initial state and the final state after one whole revolution ($t = \SI{100}{\second}$) for the Exner pressure variable. Grid resolution varies from $15\times 15 \times 15$ to $100 \times 100 \times 100$. The dashed line displays second-order convergence.

\subsection{Baroclinic vortex}

The second test case repeats the comparison for a stationary baroclinic vortex. We consider a stratified background atmosphere with a stationary, axisymmetric Gaussian Exner-pressure anomaly. The velocity, density, and potential-temperature perturbation fields are derived analytically from geostrophic and hydrostatic balance relations. In order to achieve numerical balance from the analytical balance, we run a QG projection to remove the numerical imbalance. For the fully compressible configuration, a pseudo-incompressible projection follows the QG step to suppress acoustic waves.

The vortex is centered at $\mathbf{x}_c = (0, 0, 0)$ within the computational domain
\[
    \Omega = [-1.0, 1.0]^2 \times [-0.1, 0.1]\ \si{\kilo\meter}.
\]
The Exner-pressure anomaly is specified as
\begin{align}
    \pi'(x,y,z) = \pi_{\mathrm{a}}
    \exp\!\left(-\frac{r^2}{R^{2}}\right)
    \left[ 1 - \left(\frac{\delta z}{H}\right)^{2} \right]^{3},
\end{align}
where $r^2 = (x-x_c)^{2} + (y-y_c)^{2}$ and $\delta z = z-z_c$. The perturbation is defined within the region $\{ (r, z) \mid r < R, \delta z < H \}$ and vanishes elsewhere. The parameters are set to $\pi_{\mathrm{a}} = 10^{-5}$, horizontal length scale $R = \SI{2}{\kilo\meter}$, and vertical scale $H = \SI{400}{\meter}$.
Since the velocity fields are initialized using the geostrophic balance equation, we set the Coriolis parameter to $f = \SI{0.1}{\second^{-1}}$ to simulate small Rossby number and keep the full compressible run from collapsing. In addition, we prescribe a stable stratification with a Brunt–Väisälä frequency of $N^2 = \SI{e-3}{\second^{-2}}$.

\begin{figure}[htbp]
    \centering
    \includegraphics[width=0.73\linewidth]{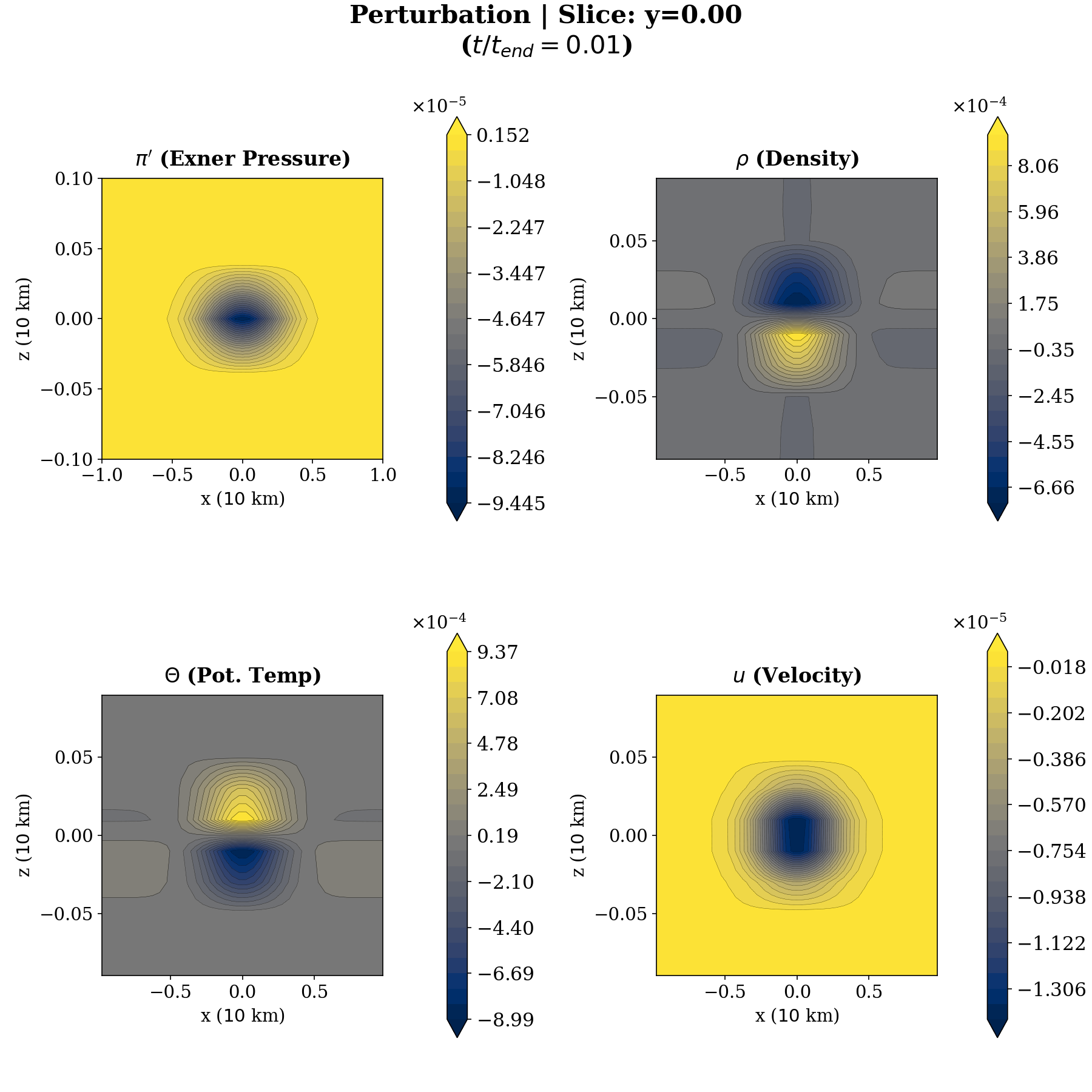}
    \caption{Initial numerically balanced baroclinic vortex at $y = 0$ after QG projection. Panels show perturbations for Exner-pressure $\pi'$, density $\rho'$, potential-temperature $\Theta'$, and horizontal velocity $u$.}
    \label{fig:baroclinic-initial}
\end{figure}

Figure~\ref{fig:baroclinic-initial} shows the vertical slice of the numerically balanced initial state at $y=0$ after the QG projection. The contours illustrate the Exner-pressure $\pi'$, density $\rho'$, potential-temperature $\Theta'$, and horizontal velocity $u$ perturbations.

\begin{figure}[htbp]
    \centering
        \includegraphics[width=1\linewidth]{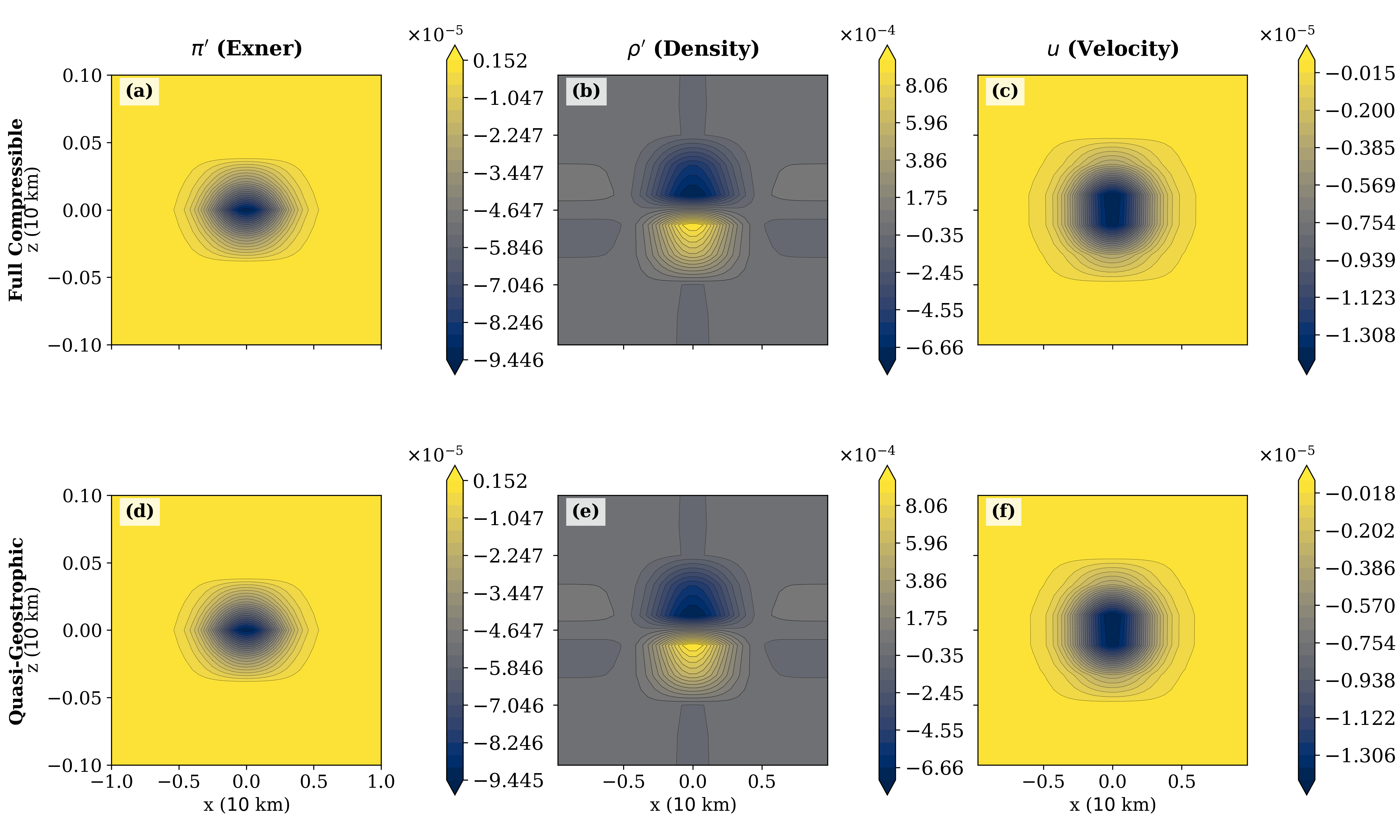}
        \caption{Vertical slice ($y=0$) of the solution at $t = \SI{100}{\second}$ for (top) full compressible and (bottom) quasi-geostrophic runs.}
        \label{fig: baroclinic-end-state}
\end{figure}

\begin{figure}[htbp]
    \centering
        \includegraphics[width=1\linewidth]{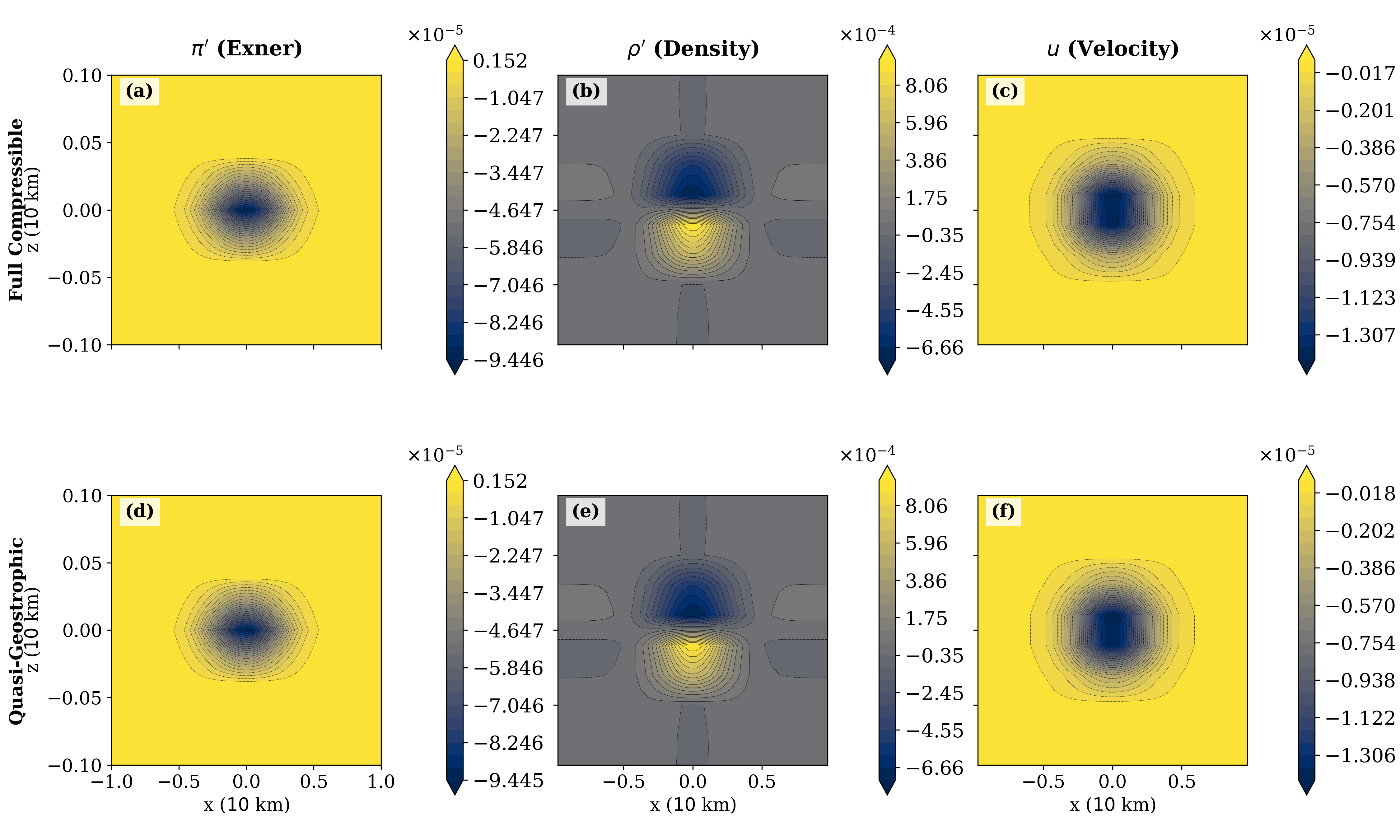}
        \caption{Vertical slice ($y=0$) of the solution at $t = \SI{300}{\second}$ for (top) full compressible and (bottom) quasi-geostrophic runs.}
        \label{fig: baroclinic-end-state-long}
\end{figure}

For the comparison, we integrate both the fully compressible and QG configurations for $t = \SI{100}{\second}$ and $t = \SI{300}{\second}$. Figure~\ref{fig:  baroclinic-end-state} compares the resulting solutions for $t = \SI{100}{\second}$. The fully compressible run (Fig.~\ref{fig: baroclinic-end-state}, top row) shows weak deviations from the initial state as a result of the internal waves. On the other hand, the QG configuration (Fig.~\ref{fig: baroclinic-end-state}, bottom row) maintains the balanced structure and deviates minimally from the initial state. Figure~\ref{fig: baroclinic-end-state-long} shows the simulation run for $t = \SI{300}{\second}$. One can see that the vortex structure does not change noticeably from the shorter simulation run in both QG and fully compressible runs.

\begin{figure}[htbp]
    \centering
    \begin{subfigure}[b]{0.9\textwidth}
        \centering
        \includegraphics[width=\linewidth]{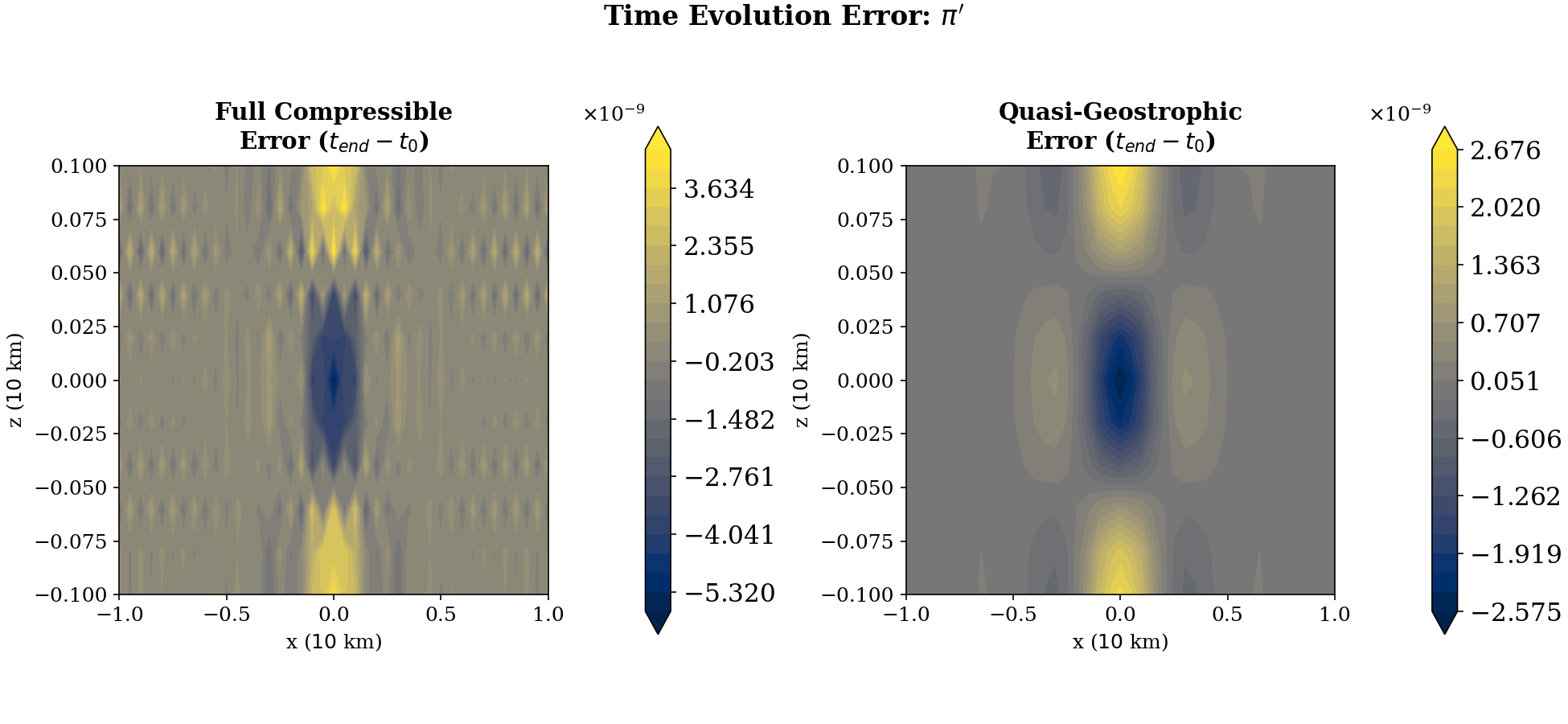}
        \caption{Exner-pressure $\pi'$ error}
        \label{fig:err-pressure}
    \end{subfigure}
    
    \vspace{1em} 
    
    \begin{subfigure}[b]{0.9\textwidth}
        \centering
        \includegraphics[width=\linewidth]{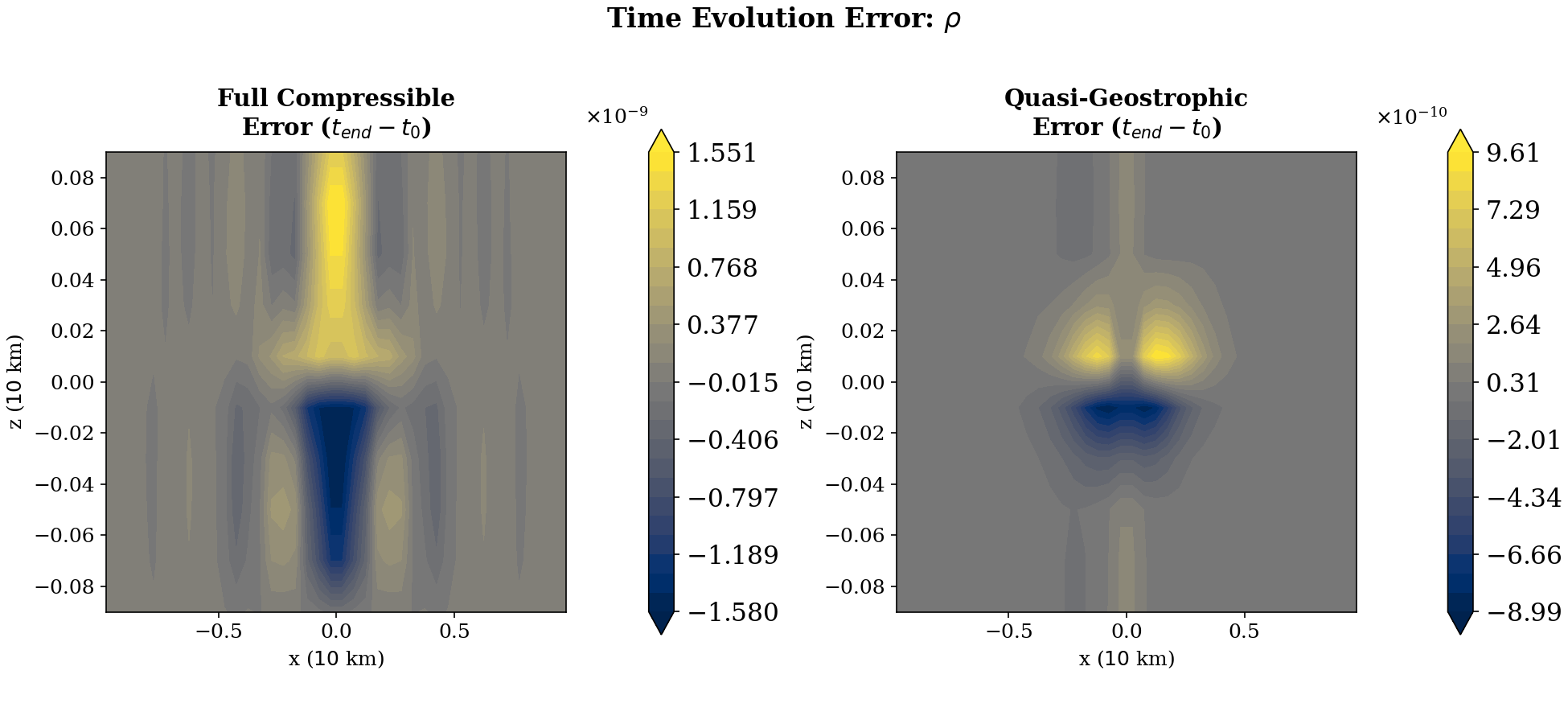}
        \caption{Density $\rho$ error}
        \label{fig:err-density}
    \end{subfigure}

        \begin{subfigure}[b]{0.9\textwidth}
        \centering
        \includegraphics[width=\linewidth]{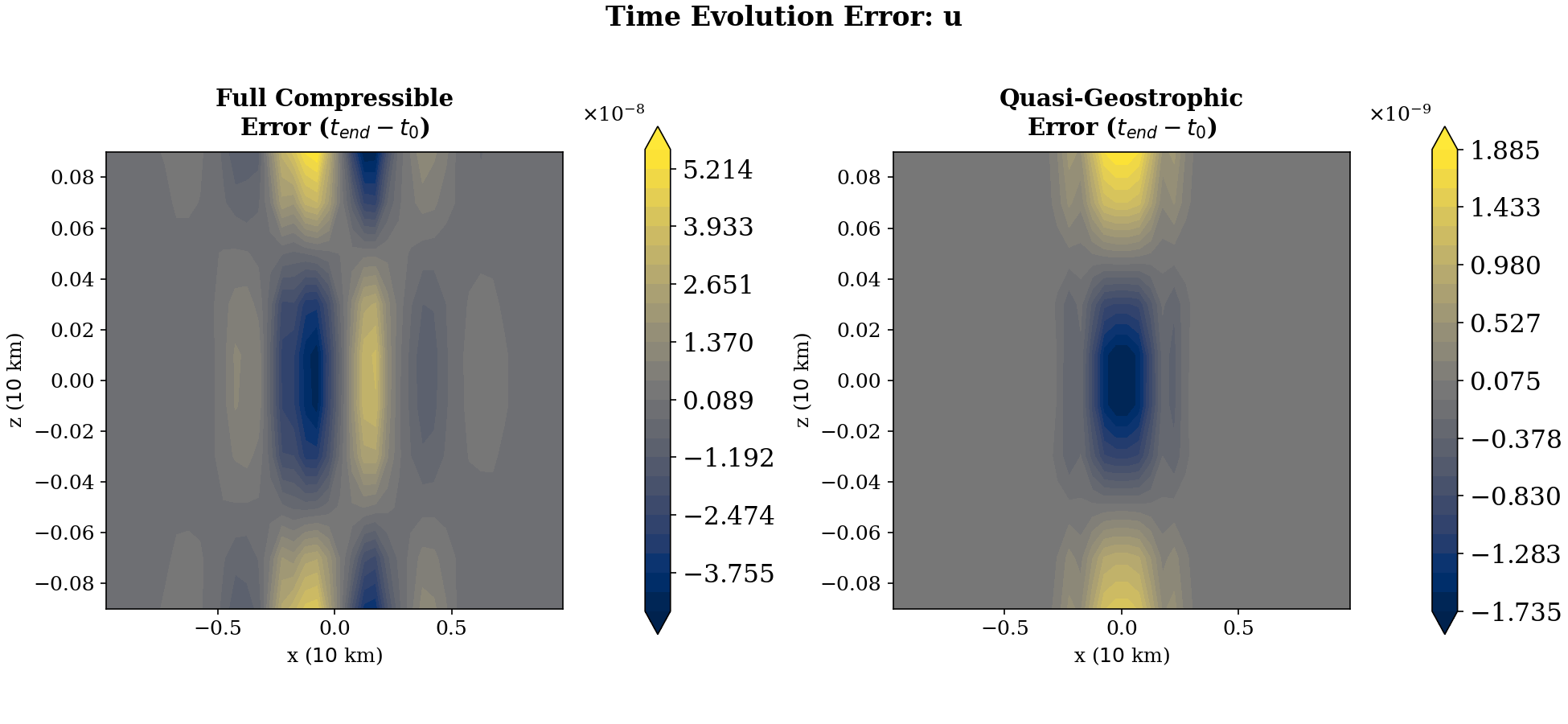}
        \caption{Density $\rho$ error}
        \label{fig:err-velocity}
    \end{subfigure}
    
    \caption{Difference between initial and final states ($t=\SI{100}{\second}$) for (a) Exner-pressure perturbation, (b) density and (c) horizontal velocity.}
    \label{fig:baroclinic-error}
\end{figure}

\begin{figure}[htbp]
    \centering
    \begin{subfigure}[b]{0.9\textwidth}
        \centering
        \includegraphics[width=\linewidth]{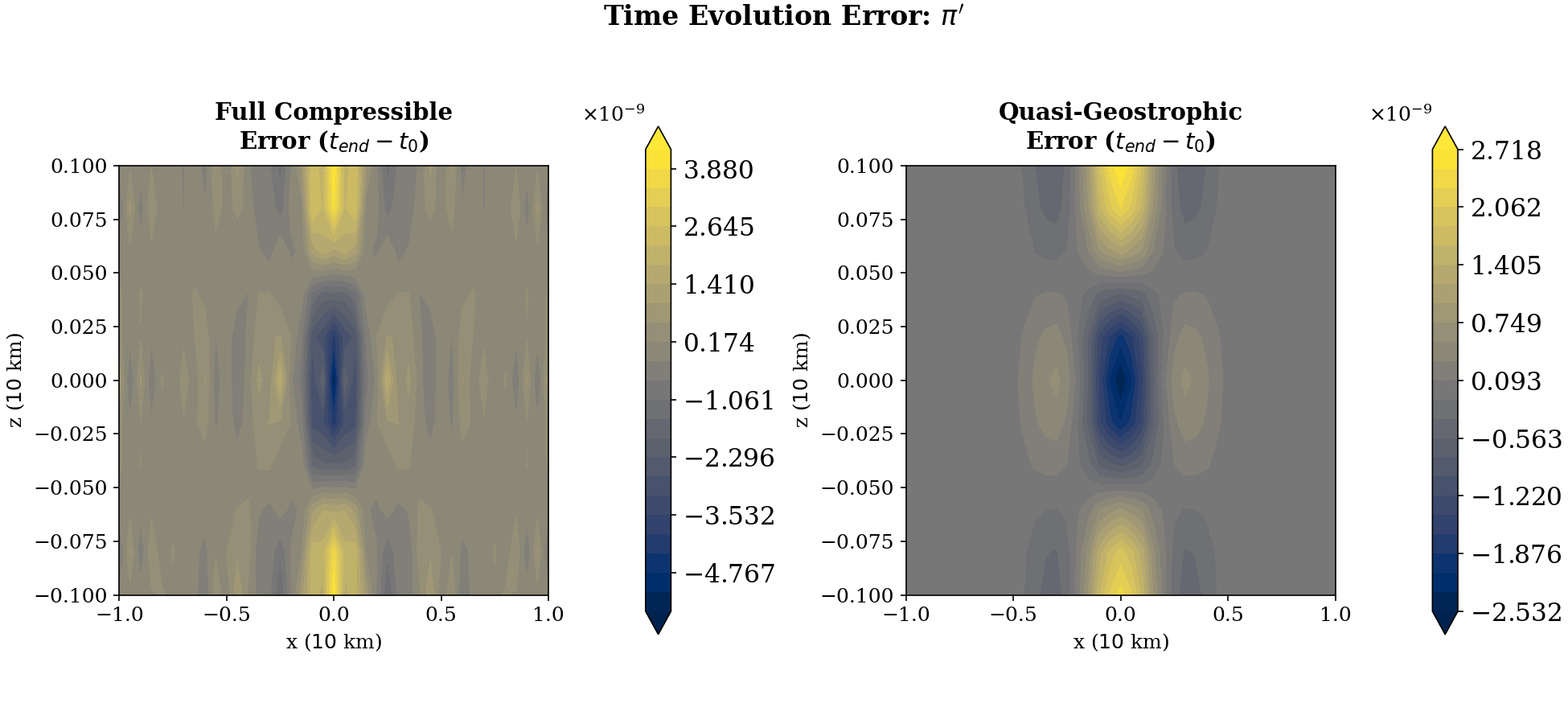}
        \caption{Exner-pressure $\pi'$ error}
        \label{fig:err-pressure-long}
    \end{subfigure}
    
    \vspace{1em} 
    
    \begin{subfigure}[b]{0.9\textwidth}
        \centering
        \includegraphics[width=\linewidth]{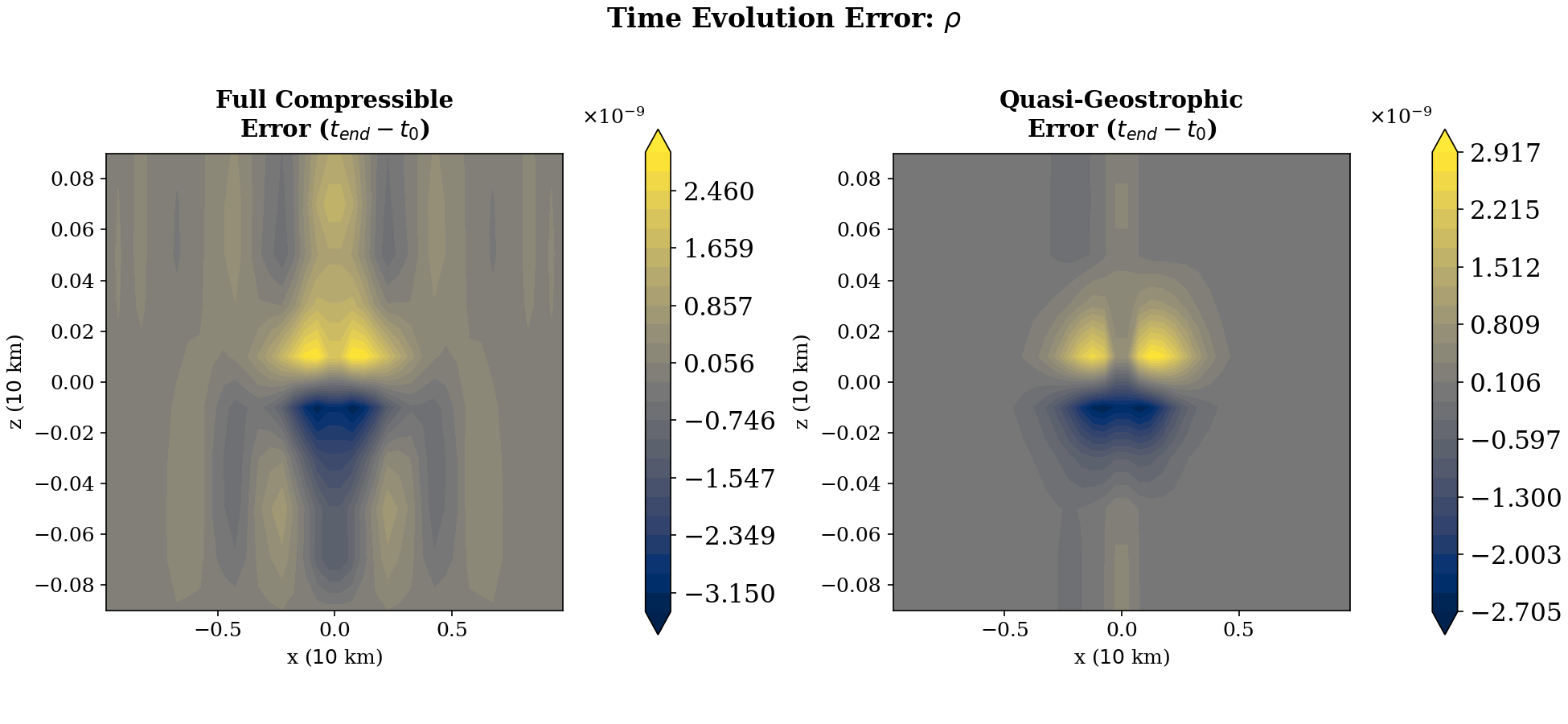}
        \caption{Density $\rho$ error}
        \label{fig:err-density-long}
    \end{subfigure}

        \begin{subfigure}[b]{0.9\textwidth}
        \centering
        \includegraphics[width=\linewidth]{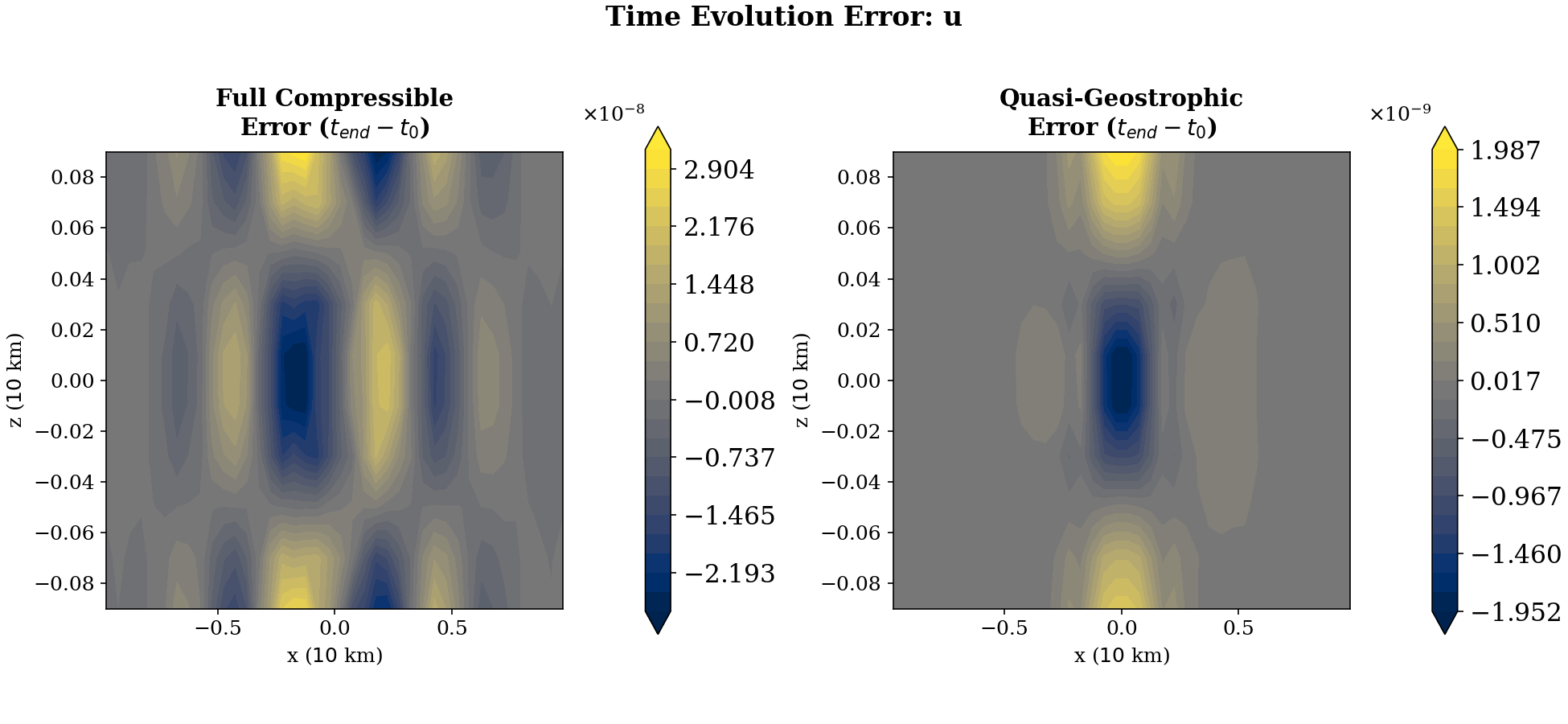}
        \caption{Density $\rho$ error}
        \label{fig:err-velocity-long}
    \end{subfigure}
    
    \caption{Difference between initial and final states ($t=\SI{300}{\second}$) for (a) Exner-pressure perturbation, (b) density and (c) horizontal velocity.}
    \label{fig:baroclinic-error-long}
\end{figure}

Finally, we quantify the deviation from the balanced state by computing the difference between the initial and final states ($t=\SI{100}{\second}$ and $t=\SI{300}{\second}$). Figure~\ref{fig:baroclinic-error} presents these deviations for Exner-pressure and density. One can notice the existence of internal waves in the full compressible run, which are fully suppressed in the QG run. The deviation in the QG run is smaller, as expected. Figure~\ref{fig:baroclinic-error-long} shows the results of the simulation run for $t =\SI{300}{\second}$. For a longer simulation, the waves and the errors are stronger, and the QG run has remained more stable than the full compressible run.

\section{Discussion and conclusions}
\label{sec: discussions-and-conclusions}

The aim of this work was to fill the gap in the model proposed in BK19 by creating access to the quasi-geostrophic regime from within the same compressible, semi-implicit solver that already provided access to soundproof and hydrostatic limits. Starting from the scale-dependent asymptotic framework of \citet{Klein2022QG-DL-Ekman:Framework}, we showed that quasi-geostrophic regime can be derived from the BK19 implicit step by utilizing the same method of asymptotic analysis used in the continuous theory. This way, the QG system appears as the direct result of the distinguished limit of the Rossby number on the existing discretization without needing to compute the potential vorticity externally.

In order to achieve the equivalence between the numerical asymptotic model and the continuous one, we first define the discrete curl operator consistent with the BK19 discretization, ensuring compatibility with the discrete gradient and divergence operators defined therein. The curl operator provides the groundwork to define both the numerical and the physical vorticity quantities, which, on the one hand, appear in the leading-order QG elliptic pressure equation and, on the other hand, are shown to remain the prognostic variable throughout the simulation. The asymptotic analysis of the equations in the implicit step then produces the QG Poisson-type equation for the leading-order Exner pressure perturbation that mirrors the QG Poisson equation in the continuous theory. This analysis guarantees that the switching between full compressible and the QG regimes is, in fact, based on a correspondence between the numerical leading-order dynamics and the continuous ones.

From the algorithmic point of view, we have introduced two new switches, $\alpha_H$ and $\alpha_u$, that provide control over the elliptic problem and the horizontal-momentum update. The switch $\alpha_H$ determines whether the elliptic equation solved in the implicit step is the full Helmholtz-type problem associated with full compressible, or the reduced QG Poisson problem. The switch $\alpha_u$ controls whether the horizontal momenta are updated prognostically or diagnosed from geostrophic balance. Together with the existing pseudo-incompressible and hydrostatic switches and the new access to the quasi-geostrophic, BK19 now offers access to four main regimes in the atmospheric dynamics within a single framework: full compressible, pseudo-incompressible, hydrostatic, and quasi-geostrophic.

These analytical results are confirmed by the numerical experiments shown in Section \ref{sec: numerical-results}. In the barotropic vortex test, the QG and pseudo-incompressible configurations maintain the balanced vortex over integration times for small Coriolis force, while the fully compressible configuration shows the expected collapse induced by the imbalance. The test also confirms that the rate of convergence of the scheme remains second-order in the QG mode as well.

The baroclinic vortex test further checks the behavior of the scheme in stratified, vertically sheared balanced flows. After initializing the vortex in analytic hydrostatic and geostrophic balance, a QG projection produces a numerically consistent state for a stationary vortex. At the same time, a pseudo-incompressible projection for the full compressible runs ensures that the initial states are free of acoustic waves for those runs. The result shows that the solution of the full compressible runs remains close to the QG reference, with a visible presence of internal waves, which are suppressed in the QG runs. The difference between the final state and the initial state of the fields highlights these internal waves but also shows that the remaining deviations are small. These results confirm that the discrete geostrophic switches derived in our analysis, together with the existing hydrostatic balance functionality, are consistent in practice.

Our methodology of combining continuous asymptotics directly in the numerical algorithm shows promise in capturing other regimes, such as the Obukhov scale or planetary scale dynamics in the same manner from within the numerical discretization in  future work. Additionally, further tests are needed to verify to what extent the QG regime and the full compressible regime behave consistently under a specific grid resolution and how dependent the discretization of the QG elliptic problem is on the resolution. In addition, the extension of this method to non-adiabatic flows is rather straightforward and can be formally tested in future work. The numerical algorithm of BK19 does not assume adiabaticity for the flow, and identical scale analysis can be applied to non-adiabatic flows (see e.g., \citet{Klein2010Scale-dependentFlows}, \citet{Botta1999DryAsymptotics} or \citet{Klein2000AsymptoticMethods}).

\section*{Acknowledgements}
\addcontentsline{toc}{section}{Acknowledgements}
The authors thank the  Deutsche Forschungsgemeinschaft (DFG) for funding through the Collaboration Research Center (CRC) 1114 ``Scaling cascades in complex systems'', Project Number 235221301, Project A02: ``Multiscale data and
asymptotic model assimilation for atmospheric ﬂows.''.

\appendix
\renewcommand{\thesection}{Appendix \Alph{section}}
\renewcommand{\theequation}{A.\arabic{equation}}
\setcounter{equation}{0}
\section{Discrete curl properties}
\label{app: curl-properties}

In Section~\ref{subsec: discrete-curl-operator} we defined the discrete curl. 
In this appendix we show that this discrete definition is consistent with its continuous counterpart.

Let
\begin{align}
    A &= (\vort{u})_{i+\frac{1}{2},\, j+\frac{1}{2},\, k}, \\
    B &= (\vort{u})_{i+\frac{1}{2},\, j+\frac{1}{2},\, k+1}.
\end{align}
We can expand $A$ as
\begin{align}
    A 
    &= \frac{1}{\Delta x}
       \Big(
           \Tilde{V}_{i+1,\, j+\frac{1}{2},\, k}
           - \Tilde{V}_{i,\, j+\frac{1}{2},\, k}
       \Big)
       - \frac{1}{\Delta y}
       \Big(
           \Tilde{U}_{i+\frac{1}{2},\, j+1,\, k}
           - \Tilde{U}_{i+\frac{1}{2},\, j,\, k}
       \Big) \\
    &= \frac{1}{2 \Delta x}
       \Big[
           V_{i+1,\, j+1,\, k}
           + V_{i+1,\, j,\, k}
           - V_{i,\, j+1,\, k}
           - V_{i,\, j,\, k}
       \Big] \notag \\
    &\quad
       - \frac{1}{2 \Delta y}
       \Big[
           U_{i+1,\, j+1,\, k}
           + U_{i,\, j+1,\, k}
           - U_{i+1,\, j,\, k}
           - U_{i,\, j,\, k}
       \Big].
\end{align}
The corresponding expansion for $B$ is analogous, with $k$ replaced by $k+1$ in the indices.

Therefore,
\begin{align}
    \frac{1}{2}(A+B)
    &= \frac{1}{\Delta x}
       \Big(
           \hat{V}_{i+1,\, j+\frac{1}{2},\, k+\frac{1}{2}}
           - \hat{V}_{i,\, j+\frac{1}{2},\, k+\frac{1}{2}}
       \Big)
       - \frac{1}{\Delta y}
       \Big(
           \hat{U}_{i+\frac{1}{2},\, j+1,\, k+\frac{1}{2}}
           - \hat{U}_{i+\frac{1}{2},\, j,\, k+\frac{1}{2}}
       \Big) \\
    &= (V_x - U_y)_{i+\frac{1}{2},\, j+\frac{1}{2},\, k+\frac{1}{2}}
\end{align}
where $V_x$ (and similarly $U_y$) are defined in \eqref{eq: discrete-derivative-on-cells}.

\begin{rem}
    Using this relation, one can verify discrete analogs of the standard curl identities, such as $\nabla \times \nabla \pi = 0$.
\end{rem}

\section{Obtaining physical QG Poisson equation}
\label{app: qg-poisson}

In this appendix, we explicitly obtain equation \eqref{eq: QG-inversion-equation-physical} from equation \eqref{eq: QG-inversion-equation}.

First, recall the numerical elliptic equation \eqref{eq: QG-inversion-equation},

\begin{align}
c_p \Bigl[(P\Theta)(\pi_x^{n+1})\Bigr]^{(2)}_x 
+ c_p \Bigl[(P\Theta)(\pi_y^{n+1})\Bigr]^{(2)}_y
+ c_p f^2\Big[\frac{(P\Theta)}{N^2}\,(\pi_z^{n+1}) \Bigr]^{(2)}_z
= f\,Z^{**}_0 - g f^2\,\Bigl[\frac{\tilde{\chi}^{**}}{N^2\chi} \Bigr]^{(2)}_z,
\end{align}

We start by reminding the reader that the leading orders of the physical pressure variable satisfy the equations \eqref{eq: zero-order-physical-hydrostatic} and \eqref{eq: first-order-physical-hydrostatic}. Notice, for $\nu$ as a placeholder for $x, y$ or $z$, we have

\begin{align}
    c_p \left[(P\Theta)\pi_{\nu} \right]_{\nu}^{(2)} = c_p \left[ (P \Theta)^{(2)} \pi_{\nu}^{(0)} + (P \Theta)^{(1)} \pi_{\nu}^{(1)} + (P \Theta)^{(0)} \pi_{\nu}^{(2)} \right]_{\nu}.
\end{align}

We know that $\pi^{(0)}$ and $\pi^{(1)}$ are functions of $z$ only, so for the case $\nu \equiv x$ or $\nu \equiv y$, the only terms remaining are 

\begin{align}
    c_p \left[(P\Theta)\pi_{x} \right]_{x}^{(2)} = c_p\left[(P \Theta)^{(0)} \pi_{x}^{(2)}\right]_x \\
    c_p \left[(P\Theta)\pi_{y} \right]_{y}^{(2)} = c_p\left[(P \Theta)^{(0)} \pi_{y}^{(2)}\right]_y
\end{align}

Therefore, the equation \eqref{eq: QG-inversion-equation} would become

\begin{align}
    &c_p\left[(P \Theta)^{(0)} \pi_{x}^{(2)}\right]_x + c_p \left[(P \Theta)^{(0)} \pi_{y}^{(2)}\right]_y + c_p f^2\left[\frac{1}{N^2} \left((P \Theta)^{(2)} \pi_{z}^{(0)} + (P \Theta)^{(1)} \pi_{z}^{(1)} + (P \Theta)^{(0)} \pi_{z}^{(2)}\right) \right]_z =\\
    &f P_0 \zeta_0 - gf^2 \left[\frac{1}{N^2} \left( (P \Theta)^{(2)}\chi^{\prime (0)} + (P \Theta)^{(1)}\chi^{\prime (1)} + (P \Theta)^{(0)}\chi^{\prime (2)} \right) \right]_z,
\end{align}
where, for the RHS, we used the following facts: $\chi = 1/\Theta$ and $\tilde \chi = P \chi^{\prime}$, and we dropped the $**$ superscript for simplicity.
Using equations \eqref{eq: zero-order-physical-hydrostatic} and \eqref{eq: first-order-physical-hydrostatic}, one can cancel the corresponding terms between RHS and LHS in the vertical derivative brackets to obtain:
\begin{align} \label{eq: appendix-physical-QG}
    &c_p\left[(P \Theta)^{(0)} \pi_{x}^{(2)}\right]_x + c_p \left[(P \Theta)^{(0)} \pi_{y}^{(2)}\right]_y + c_p f^2\left[\frac{1}{N^2} \left( P \Theta)^{(0)} \pi_{z}^{(2)}\right) \right]_z =
    f P_0 \zeta_0 - gf^2 \left[\frac{1}{N^2} \left( (P \Theta)^{(0)}\chi^{\prime (2)} \right) \right]_z.
\end{align}

We know from Remark (\ref{rem: background-theta-is-one}) that $(P\Theta)^{(0)} = P_0 \Theta_0 = \rho_0$. On the other hand, $\rho_0$ is a function of $z$ only, therefore, the final equation \eqref{eq: QG-inversion-equation-physical} is obtained from \eqref{eq: appendix-physical-QG}.

\printbibliography
\end{document}